# Why does time feel the way it does? Towards a principled account of temporal experience

Renzo Comolatti,[1,2*] Matteo Grasso,[2*] and Giulio Tononi[2†]

[1]University of Milan; [2]University of Wisconsin–Madison; *These authors contributed equally to this work
† Correspondence to gtononi@wisc.edu

We thank Larissa Albantakis, Isaac David, Andrew Haun, Julia Thompson, and the participants of the 2024 Qualia Structure summer school for their feedback on previous drafts. This project was made possible through support from Templeton World Charity Foundation (nos. TWCF0216 and TWCF0526). The opinions expressed in this publication are those of the authors and do not necessarily reflect the views of Templeton World Charity Foundation.

## Abstract

Time flows, or at least the time of our experience does. Can we provide an objective account of why experience, confined to the short window of the conscious present, encompasses a succession of moments that slip away from *now* to *then*—an account of why time feels flowing? Integrated Information Theory (IIT) aims to account for both the presence and quality of consciousness in objective, physical terms. Given a substrate's architecture and current state, the formalism of IIT allows one to unfold the cause–effect power of the substrate, yielding a cause–effect structure. According to IIT, this accounts in full for the presence and quality of experience, without any additional ingredients. In previous work, we showed how unfolding the cause–effect structure of non-directed grids, like those found in many posterior cortical areas, can account for the way space feels—namely, *extended*. Here we show that unfolding the cause–effect structure of directed grids can account for how time feels—namely, *flowing*. First, we argue that the conscious present is experienced as flowing because it is composed of *phenomenal distinctions* (*moments*) that are directed, and these distinctions are related in a way that satisfies *directed inclusion*, *connection*, and *fusion*. We then show that directed grids, which we conjecture constitute the substrate of temporal experience, yield a cause–effect structure that accounts for these and other properties of temporal experience. In this account, the experienced present does not correspond to a process unrolling in "clock time," but to a cause–effect structure specified by a system in its current state: time is a structure, not a process. We conclude by outlining similarities and differences between the experience of time and space, and some implications for the neuroscience, psychophysics, and philosophy of time.

# 1. Introduction

Why does an experience feel the way it does? On a morning walk, you hear the notes of the bird's song succeed one another and see it fly across the blue expanse of the sky. Can we provide a scientific account for the quality of consciousness, including the feeling of why time feels flowing, space feels extended, and the sky feels blue?

Integrated Information Theory (IIT; Albantakis et al., 2023; Oizumi et al., 2014; Tononi, 2004, 2008) aims to do just this: to provide a principled and comprehensive account of phenomenal properties in physical terms. First, it identifies the essential properties of consciousness—those that are true of every conceivable experience. It then formulates these properties in physical, operational terms. In doing so, IIT provides the tools to identify the substrate of consciousness (a *complex*) and unfold its cause–effect power (the *cause–effect structure* it specifies, composed of *causal distinctions* and *relations*). According to IIT, the cause–effect structure specified by a given substrate in its current state is sufficient—without additional ingredients—to fully account for the quality (content) and quantity of experience. In a previous paper, we showed how the cause–effect structures specified by undirected grids can account for the feeling of extendedness that characterizes spatial experiences (Haun & Tononi, 2019). In this paper, we employ the formalism of IIT to account for why time feels flowing.

The time under inquiry here is not what is measured by clocks ("clock time") but the subjective time of experience, the feeling of a short window of a conscious "present," composed of moments that succeed one another, in which we feel something is happening "now," something happened "then," and often that something will come "next."

We start from the basic phenomenology of time and characterize its fundamental properties: a temporal experience is a kind of phenomenal structure called a *phenomenal flow*, composed of distinctions and relations characterized by *directedness*. We then propose an account of phenomenal flow in physical terms, where "physical" is understood in a purely operational sense (manipulations and observations on a substrate), yielding a transition probability matrix (TPM). Specifically, we show that a certain kind of substrate—namely, a directed grid—specifies a cause–effect structure that can account for the phenomenal properties of temporal experience.

# 2. Phenomenology of time

Like the feeling of spatial extendedness, the feeling of temporal flow is not only pervasive in our experience but also partially penetrable. In other words, unlike, say, the feeling of color or pain, we can partially dissect the basic structure of phenomenal time through introspection, even though its fleeting nature makes it more difficult to characterize than phenomenal space (Augustine, 2009; Haun & Tononi, 2019; Husserl, 1991).

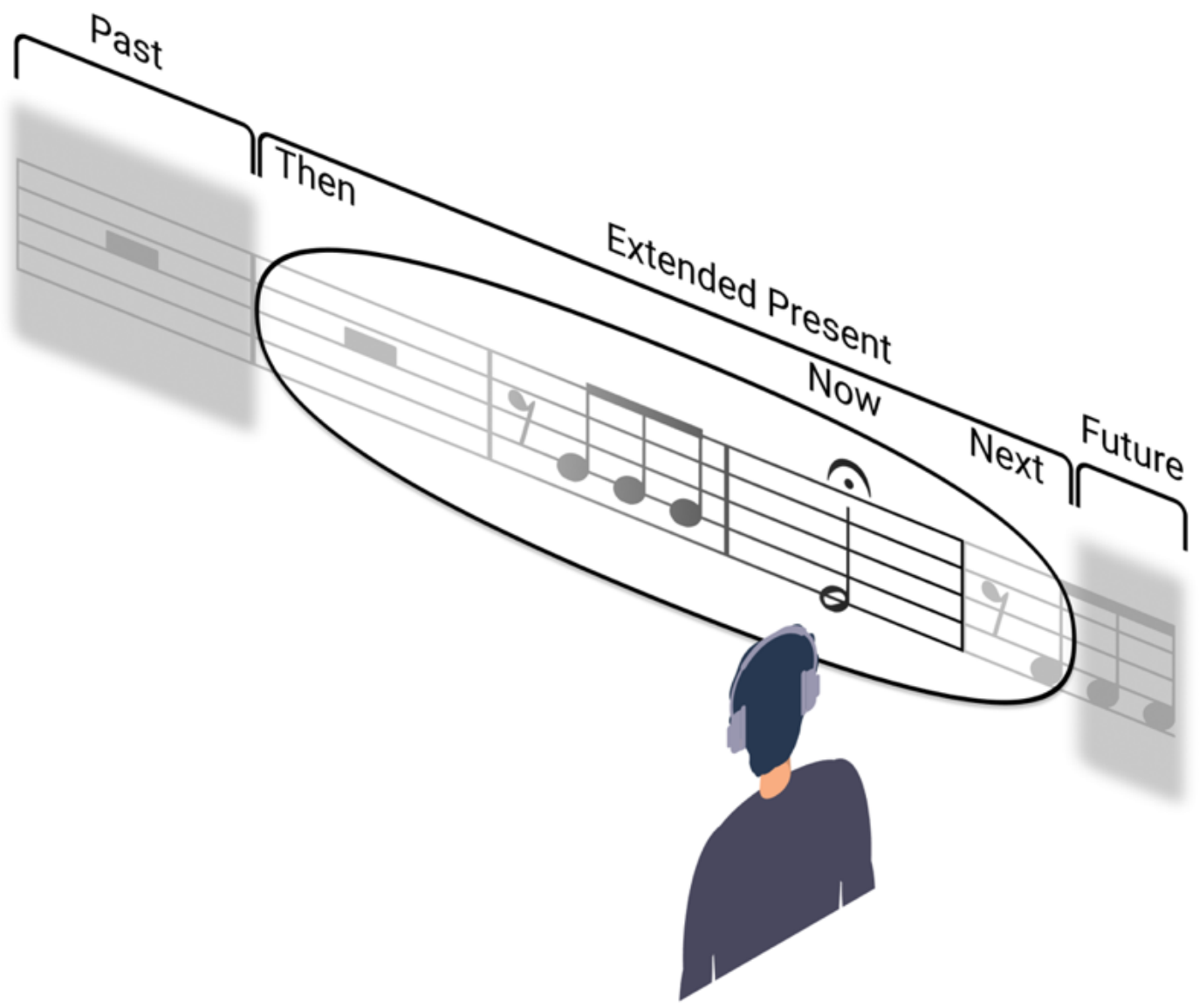

**Figure 1:** A depiction of temporal experience. The bubble indicates the content of a single experience, represented for convenience by a musical score—a few bars of Beethoven's Symphony no. 5. The experience can be triggered by clicking the play button at this page until the note E is perceived (the fourth note). The *present* is experienced as *extended*—as having a duration (e.g., the four notes played). Moments within the extended present (associated with notes, pauses, and their combinations) flow from the *now* (e.g., the E note that is playing) towards the *then* (e.g., the three G notes just played). The silence before the first note (grayed out on the left) has vanished from experience into what we call the *past*, although it may be summoned within experience by recalling it. The extended present may include a feeling of what will come *next* (the upcoming notes). Beyond that lies what we call the *future*.

Below, we highlight some fundamental features of phenomenal time that we intend to account for. Consider the temporal phenomenology of hearing a melody—say, the first few notes of Beethoven's Fifth Symphony (Figure 1), abstracting away from the phenomenal qualities of sound (we might be listening to another melody, or to speech, or even to silence, and phenomenal time would still be flowing).

First, our experience comprises an *extended present*, structured by phenomenal distinctions, called *moments*, which are related in a special way. The present is extended in the sense that we hear the melody, rather than a single note: our experience contains the note we heard just *now* together with a few other notes we heard just *then* (moments ago, but still *present* in our consciousness), and may contain the feeling of what will come *next* (the note that we will hear in a moment, already present in our experience, albeit less vividly). On the other hand, we do not experience notes outside the present, whether in the future, beyond the next (Figure 1, right), or in the past, beyond the then (Figure 1, left). The notion of the extended or "specious" present was popularized by William James as "the short duration of which we are immediately and incessantly sensible" (James, 1890) , borrowing from E. Robert Kelly (Clay, 1882).

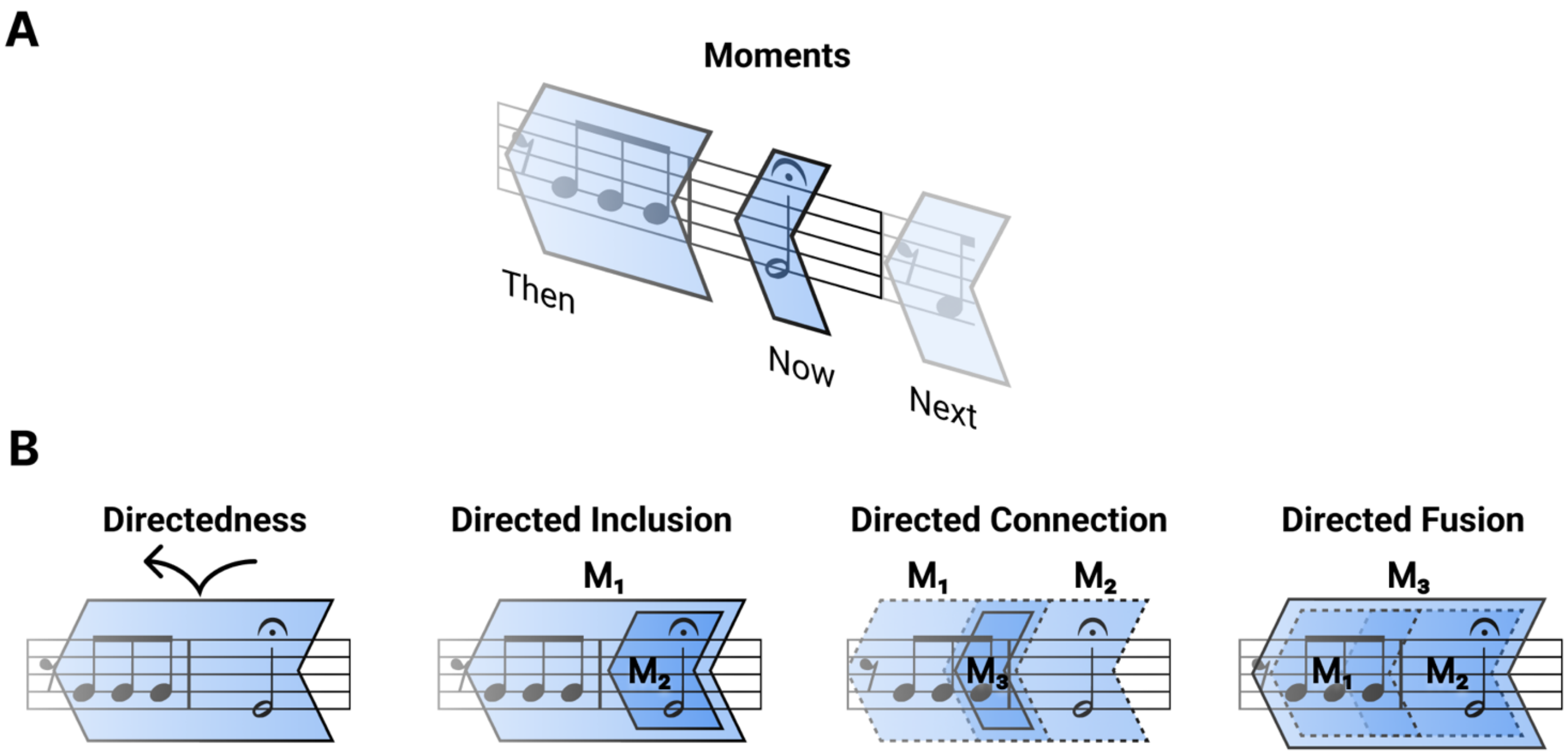

**Figure 2:** Phenomenology of temporal experience. (A) The distinctions composing the phenomenal structure of temporal experience are called *moments* (blue block arrows). (B) The fundamental properties of temporal experience are *directedness* and *directed inclusion*, *connection*, and *fusion*: (i) Moments are directed; they point away from themselves. (ii) Moments include and are included by other moments and do so in a directed manner, either forward (towards the now, in the case depicted) or backward (towards the then). (iii) Moments connect when they partially overlap each other in a directed way, such that one is the predecessor ($M_1$) and the other the successor ($M_2$), and their overlap is also a moment ($M_3$). (iv) Moments that connect also fuse with one another such that their union is also a moment that includes them in a directed way and nothing else.

Second, the present is structured by phenomenal relations that make it feel *directed*, yielding a temporal *flow*: Within the present, we experience moments, such as those corresponding to individual notes or pauses, that appear to "flee away," *directed* towards what we call the past (the flow of experienced time thus runs "backward"—from the now toward the past—rather than "forward"—from the now toward the future—like clock time). As characterized here, the feeling of temporal flow only refers to the experience of moments "fleeing away," rather than to the experienced "speed" of their flow (see Discussion).

The moments that compose the present are bound together by directed relations that order them, as we shall argue, according to *inclusion*, *connection*, and *fusion*. These relations yield an *experience of succession*, rather than a *succession of experiences* (James, 1890).

Since the experience of the extended present does not necessarily include the feeling of what will come next, our account will mainly focus on the flow of time from the now to the then—what in the philosophical literature has been referred to as "retention" (Husserl, 1991). Nonetheless, in the Discussion we will show how this approach can also account for the experience of "protention"—occasionally extending to what we feel will come next.

## 2.1. *Moments*

Let us dissect the phenomenology of temporal flow in more detail and introduce some nomenclature (Figure **2**A). As already mentioned, the phenomenal distinctions that compose phenomenal time are called *moments*. Moments can be as short as an "instant" (the shortest moment one can phenomenally resolve), as long as the entire present (the "total" moment), or anything in between. For instance, in the musical example, a moment may encompass a single note, a single pause, a note and a pause, two notes separated by a pause, and so on. Further, moments can be close to the conscious *now*, to the conscious *then*, or anytime in between. There are various estimates about the duration of the conscious present, from a few hundred milliseconds up to three seconds of clock time (Dainton, 2023; Pöppel, 2009), and of the grain of a conscious instant, typically a few tens of milliseconds (Herzog et al., 2016; White, 2018). Moments are fleeting, whether short and long, now and then, but owing to the *relations* that bind them together, they compose the flow of experienced time within the present. In what follows, we argue that four fundamental properties, characterizing moments and their relations, are necessary and sufficient for the experience of time: *directedness*, *directed inclusion*, *directed connection*, and *directed fusion* (Figure **2**B). These properties apply to all moments, except for the moment corresponding to the entire present (the "total moment" and the moments corresponding to individual instants.

## 2.2. *Directedness*

Moments are directed, each of them pointing away from itself—fleeing away from the now and towards the then. For this reason, we represent the phenomenology of moments in time by arrows (pointing away from the now towards the then).

## 2.3. *Directed inclusion*

For any moment, there are always other moments which include it or are included by it. By virtue of being directed, inclusion can be of two kinds: moments can be included towards the now (*forward inclusion*) or towards the then (*backward inclusion*). Forward inclusion is such that the included moment is a subset of the including moment but is aligned on the latest instant on which they both overlap: they "share their ending"—the last instant towards the now they both include. Similarly, backward inclusion is such that the included moment is a subset of the including moment and is aligned on the earliest instant on which they both overlap: they "share their beginning"—the first instant towards the then they both include. Directed inclusion captures the fact that every moment feels nested within the structure of the present, encompassing a certain period (determined by the moments it includes) and having a temporal location within the present (determined by the moments that include it).

## 2.4. *Directed connection*

For any moment we can always find *predecessor* moments that overlap it partially and asymmetrically towards the then, and *successor* moments that overlap it partially and asymmetrically towards the now. Directed connection is asymmetric because there is an intrinsic ordering within *phenomenal* time: a moment that is connected to a successor moment cannot be its successor but only its predecessor, and a moment connected to its predecessor cannot be its predecessor, only its successor. When two moments overlap, there is always another moment that covers exactly their overlap—the *connecting* moment (or *connection*). This third moment is included by both in a directed way, such that the connecting moment is *forward-included* by the predecessor and *backward-included* by the successor. For any two overlapping moments, one can always find a moment they connect onto (*directed connection down*). Moreover, every moment is also the connection of two overlapping moments, such that it is included by both of them and covers their overlap (*directed connection up*). Directed connection accounts for the directed ordering of moments within the present according to relations of *succession* and *predecession*.

## 2.5. *Directed fusion*

For any moment, one can always find another connected moment with which it *fuses*, such that together they compose a third moment that includes both of them in a directed way (either *backward* or *forward*) and coincides with their union (*directed fusion up*). Every moment is also the fusion of two connected moments (*directed fusion down*), one towards the now and the other towards the then, such that it includes both of them and coincides with their union. Fusion accounts for the *fullness* of the present—that phenomenal time is not fragmented.

## 2.6. *Additional properties*

The fundamental properties of temporal flow are sufficient to derive other phenomenal properties of experienced time, such as the *period* covered by a moment, its *temporal location*, *duration*, *boundary*, and the *interval* between any two moments. These properties will be described in further detail below.

Some other phenomenal properties tightly bound to the experience of temporal flow should also be mentioned. Within the present, one or more moments can stand out because of an inhomogeneity in local properties. These are properties, such as sound or touch, that are not in themselves temporal but are typically experienced as bound to time. For instance, a sudden sound may pierce the silence (e.g., the first note in Figure **2**A), or a sudden pause interrupting a droning noise. These inhomogeneities highlight particular moments in the flowing present, without disrupting its flow, only warping it locally and often capturing our attention. But time flows in perfect silence too—say, during the expressive pauses at the end of Sibelius's fifth symphony, or throughout the provocative emptiness of John Cage's piece 4'33.

While experienced time always flows from the now to the then, another prominent phenomenal property is that we usually feel *centered* in the now (rather than in the then or in the middle of the present): when we hear a sound, it suddenly appears in the now, it stays present in our experience—fleeing away towards the then—and then disappears into the past. For example, in hearing Beethoven's Fifth Symphony, the latest note appears abruptly in the now, while the group of three notes in the previous bar is still present appear displaced towards the then (Figure **2**A). The now is the moment that is typically bound to actions: we typically feel that, when we act, we are doing so from the now rather than from the then.

# 3. Methods

We now proceed to lay out an account of the subjective properties of temporal flow in objective, operational terms, according to the principles of IIT. This means that phenomenal distinctions and relations that compose temporal flow—the moments bound by directed inclusion, directed fusion, and directed connection—must have a correspondent in the cause–effect structure specified by the substrate of temporal experience in the brain, in the causal distinctions that compose it and the way they relate.

Below, we first briefly summarize the IIT formalism (for a complete presentation see Albantakis et al. (2023) and the Integrated Information Theory Wiki (2024)). Next, we apply the formalism to unfold the cause–effect power of a directed grid—the kind of substrate that, we conjecture, can support the experience of time. All computations are performed using PyPhi (W. G. P. Mayner et al., 2018). Further details about the code and additional resources are available at https://www.iit.wiki/contents/time.

## 3.1. *Unfolding cause–effect structures*

IIT starts by applying the postulates of *existence*, *intrinsicality*, *information*, *integration*, and *exclusion* and identifying a maximum of *system integrated information* ($\varphi_s$) over the units of a substrate (Albantakis et al., 2023; Integrated Information Theory Wiki, 2024). According to IIT, a substrate of consciousness, or *complex*, must be such a maximum. Next, in line with the postulate of *composition*, the cause–effect power of the complex is unfolded in full, yielding its *cause–effect structure*. By IIT, the causal distinctions and relations that compose the cause–effect structure account for the content of the corresponding experience, with no additional ingredients. Here we focus on unfolding the cause–effect structure specified by a directed grid, assuming that the grid is part of a larger complex. Every system subset that satisfies IIT's postulates of physical existence (except for composition, which does not apply to the components themselves) specifies a causal *distinction*. A distinction consists of a *mechanism* (a subset of units in a state) that specifies a *cause purview* and an *effect purview* (each a subset of units in a state). Overlaps among causes and/or effects of one or more distinctions specify *causal relations* (Figure 3A, top left).

### 3.1.1. *Distinctions*

A mechanism specifies a causal distinction if it satisfies IIT postulates in that (i) it has cause–effect power (existence postulate), that is, it can take and make a difference with respect to itself or other units; (ii) it has cause–effect power within the system (intrinsicality postulate); (iii) its cause–effect power is specific (information postulate), that is, being in its specific current state, it selects a state for its purviews (the one with maximal *intrinsic information* $\boldsymbol{ii}$ on the input side for cause purview and on the output side for effect purview), and this state is congruent with the *cause–effect state* selected by the complex as a whole; (iv) its cause–effect power is irreducible (integration postulate), that is, the distinction's *integrated information* ($\varphi_d$, the minimum amount of $\boldsymbol{ii}$ lost by partitioning mechanism and purview) is positive; and (v) the amount of $\varphi_d$ it specifies is maximal across other candidate purviews (exclusion postulate). In sum, a causal distinction comprises a mechanism linking a cause purview and an effect purview, and has an associated $\varphi_d$ value.

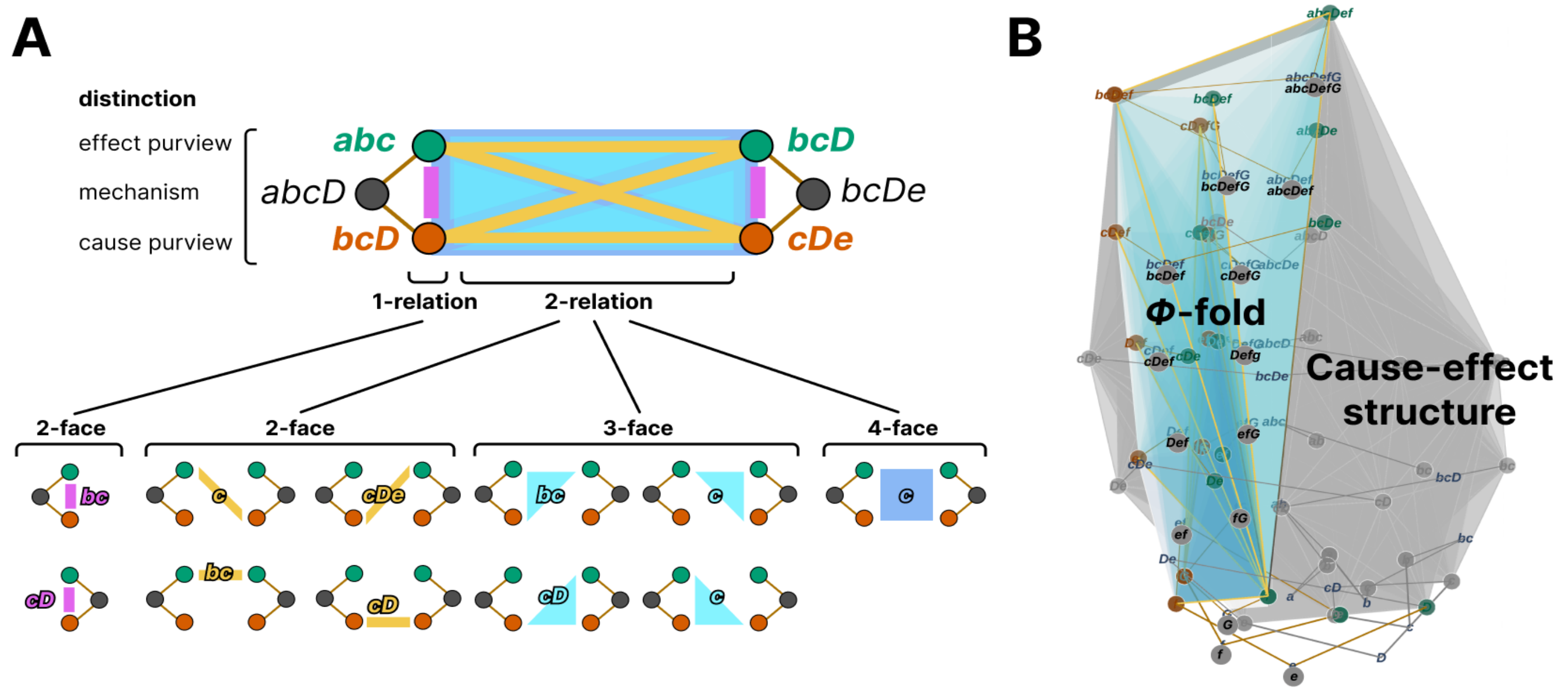

**Figure 3:** A cause–effect structure and its components. (A) A cause–effect structure is composed of causal distinctions and relations. Distinctions are specified by irreducible mechanisms (subsets of the substrate's units, indicated by black circles) linking a cause and an effect purview (red and green circles, respectively) over subsets of the substrate's units. Causal relations occur when one or more distinctions overlap congruently (same unit state) over two or more purviews. Each purview overlap in a relation is called a relation face and can involve causes, effects or both. Each causal relation is thus a bundle of relation faces. The degree of a relation is the number of distinctions involved in the overlap, while the degree of a relation face is the number of purviews contributing to it. In the example, the causal relations involving two distinctions *abcD* (left) and *bcDe* (right) and their respective mechanism (black circle), cause (red circle) and effect (green circle) purviews, are depicted. The binary states of the mechanism and purviews is represented by -1 (OFF) in lowercase and +1 (ON) in uppercase. Below, the relation faces composing the 1- and 2-relations are shown with their respective overlap (i.e. units in a state). Second-degree faces (or 2-faces) are depicted as edges (yellow and magenta), and higher-degree faces are depicted as surfaces (blue). Shown are two distinctions, with their 1- and 2-relations and the 2-, 3- and 4-relation faces they could have. (B) Causal distinctions and relations compose a cause–effect structure (depicted in gray), from which sub-structures (or *Φ*-folds, in blue) can be isolated. For a full description of distinctions, relations, and cause–effect structures and how to compute them using the formalism of IIT 4.0, see Albantakis et al. (2023) and the "Integrated Information Wiki" (2024).

### *3.1.2. Relations*

Causal relations capture the way in which causal distinctions are bound together within a cause–effect structure. There is a relation if cause and effect purviews overlap congruently (i.e., they specify the same state) over a subset of their units (Figure 3A). The purviews specified by a set of distinctions can overlap in different ways—depending on whether the overlap involves causes, effects, or both, and on the number of purviews that overlap. Each of the purview overlaps in a relation is called a *face*. The units in the overlap constitute the *face purview*, and the union of the face purviews constitutes the *relation purview*. The relation irreducibility value ($\varphi_r$) is calculated by unbinding one distinction at a time and finding the one that makes the least difference. This is calculated by multiplying the average $\varphi_d$ per unique purview unit by the size of the overlap across all faces (the number of units in the relation purview) and taking the minimum value across distinctions in the relation.

A relation that binds $n$ distinctions is called an $n^{\text{th}}$-degree relation (or $n$-relation for short) and a face that binds $k$ purviews within a relation is called a $k^{\text{th}}$-degree face (or $k$-face for short). In the case of relations involving two distinctions $\boldsymbol{d} = \{d_1, d_2\}$, each with a cause and an effect purview, we have a set of four purviews $z_d = \{e_1, c_1, e_2, c_2\}$ (where $e$ and $c$ stand for effect and cause purview, respectively; see Figure 3A for an example).There are nine potential relation faces f($z_d$) across the two distinctions: one 4-face involving all four purviews; four 3-faces involving three purviews (either two effects and one cause, or two causes and one effect); and four 2-faces involving two purviews (either a cause and an effect, two causes, or two effects). Finally, there are two potential 1-relations (a self-relation between the cause and effect of each distinction).

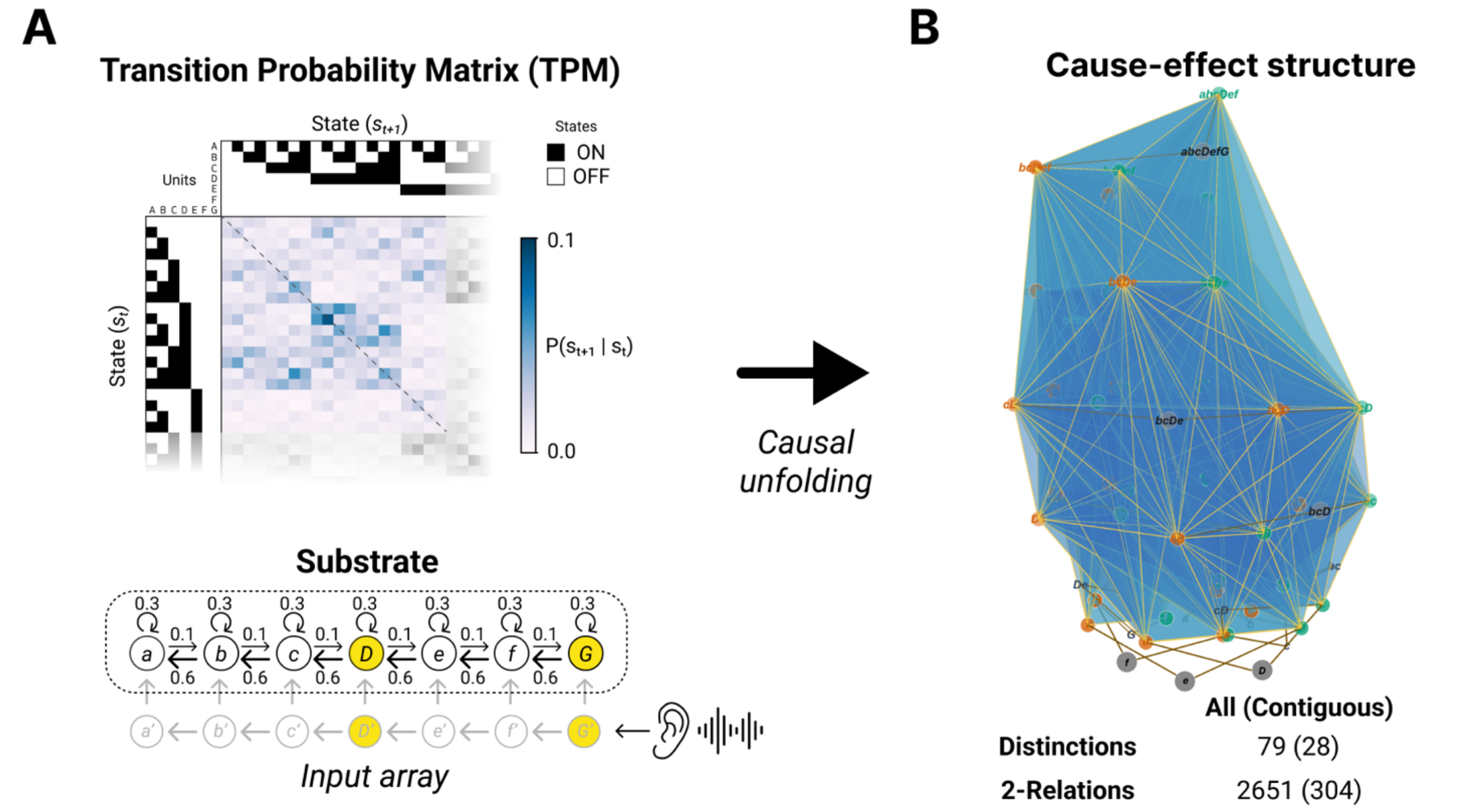

**Figure 4:** Substrate model of a directed 1D grid and its cause–effect structure. (A) Below, the substrate consisting of a directed 1D grid with seven probabilistic units (*abcDefG*), where the binary state is represented by -1 (OFF) in lowercase and +1 (ON) in uppercase. Each unit is characterized by a self-connection weight of $w$ = 0.3, an outgoing lateral connection with a weight of $w$ = 0.6 to one neighboring unit, and with a lesser weight of $w$ = 0.1 to the other neighbor. The substrate is assumed to be part of a larger complex. Outside the complex is an input array conveying sensory input. The input array functions as a delay line, percolating activations from the ear from unit $G'$ to $a'$. The input array also drives the activation state of the directed grid, which "endorses" its driven state through its self-and lateral connections that undergo short-term plasticity. Above, the associated transition probability matrix (TPM) which contains all information needed to unfold the cause–effect structure of the substrate model. Each state $s_t$ (rows) can transition to a state $s_{t+1}$ (columns) with probability $P(s_{t+1} \mid s_t)$. The binary states are represented as blocks (+1 as black, -1 as white) and only the first twenty states are shown. (B) Unfolded cause–effect structure of the seven-unit directed grid. Each distinction consists of a mechanism (black units) linked by brown lines to its cause (red units) and its effect (green units). Lower-order distinctions are depicted towards the bottom and higher-order distinctions towards the top. Only 1st- and 2nd-degree relations are plotted, with 2nd-degree faces depicted as edges (yellow) and higher-degree faces depicted as surfaces (blue).

Together, distinctions and relations compose the cause–effect structure (Figure 3B, in gray). Here we will limit our analysis to 2-relations and their underlying set of faces, in particular 2-faces, from which we can begin to characterize the cause–effect structure corresponding to temporal flow.

### *3.1.3. Contexts*

To study the contribution of individual distinctions, it is useful to decompose the cause–effect structure into sub-structures or *Φ-folds*. A relevant sub-structure is the distinction's *context*—the set of relations bound to it (Figure 3B, in blue). Through its context, a distinction is related to a set of other distinctions within the cause–effect structure. More specifically, the *purview context* is the set of relations involving a distinction's purview (either cause or effect).

In accounting for the properties of temporal phenomenology in terms of properties of the cause–effect structure, we will present the correspondence both at the "local" level of relation faces between pairs and triples of distinctions (restricting our analysis to 2-relations among them) and at the "global" level of *Φ*-folds corresponding to relation contexts.

## *3.2. Causal model of the substrate: a directed 1D grid of binary units*

The brain mechanisms and regions that support the experience of the flow of time are currently not known. Here we conjecture that the experience of time is supported by brain regions harboring connectivity patterns resembling directed grids. Such connectivity may be found, for instance, within the auditory cortex.

The substrate model employed in this paper is a 1D grid, assumed to be part of a larger complex, comprising seven probabilistic units *abcDefG* with binary state (-1, or OFF, indicated with lowercase, and +1, or ON, indicated with uppercase; Figure 4A). This is considered a "macro" state, corresponding to an interval of the order of 30 milliseconds or so of clock time (see below). Each unit has a self-connection (weight of $w = 0.3$), a stronger outgoing lateral connection ($w = 0.6$) to one of its two neighboring units, and a weaker outgoing lateral connection ($w = 0.1$) to the other neighboring unit. Each unit also receives a feedforward input from a sensory interface (*input array*; Figure 4A, bottom), assumed to be outside the complex, also comprising seven units. The input array not only provides bottom-up inputs that drive the activation of the units of the 1D grid; is also works as a delay line, such that activation percolates sequentially from unit $G'$ to unit $a'$.

The 1D grid constituted of units *abcDefG* does not percolate activity patterns on its own, but "endorses" the activity macro state driven by the input array through an activation function that is the combination of two sub-functions (see W. Mayner et al., 2024) for details on their implementation). The first function $f_1(x_k,s_k)$ assures that grid units are reliably turned ON and OFF if the feedforward sensory input is ON and OFF, respectively. If the unit's current state $s_k$ differs from the sensory driving input $x_k$, the unit's state flips. The second function $\sigma(I^*; s_k, w_k, I)$ determines the state of each grid unit as a function of the inputs it receives through its lateral and self-connections. Each unit implements a sigmoid function of an input state $\boldsymbol{I}^*$ parametrized by the current state of the unit itself $s_k$, the connection weight $w_k$, and the current state of its input units $\boldsymbol{I}$:

$$\sigma(I^*; s_k, w_k, I) \;=\; \frac{1}{(1 + exp\;\left[-s_k \sum_{j=1}^{|I|} I_j w_{k,j} I_j^*\right]}$$

This makes connections to a unit that is ON (+1) effectively excitatory, and connections to a unit that is OFF (-1) effectively inhibitory. The state-dependent nature of this function ensures that each unit's state is endorsed by the lateral connections by adjusting the effective sign of the input to the unit (assumed to be mediated by short-term plasticity, see (W. Mayner et al., 2024)). The two functions are combined to obtain the probability of a unit turning ON by taking the one that deviates maximally from chance (i.e., the "maximally selective" one):

$$Pr(k = ON) \;=\; argmax_{p \in \{f_1(x_k,s_k), \sigma(I*;\, s_k, w_k, I)\}}\, |p \;-\; 0.5|$$

As a result, the macro state of the grid is driven by the sensory input array, while simultaneously allowing the units to endorse their current state by rapidly adjusting the strength of their intrinsic connections (at a faster time scale than that of the units' macro state).

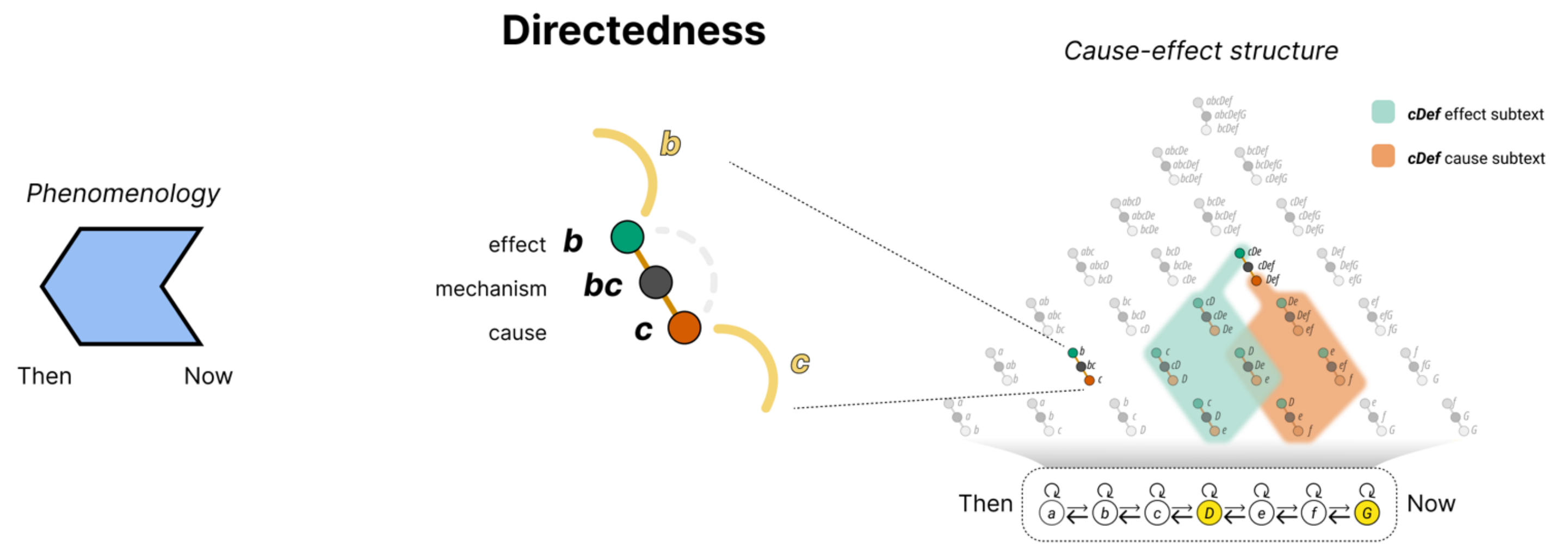

**Figure 5:** Directedness. Phenomenal distinctions in temporal experience are called moments. Moments are fundamentally directed, pointing away from themselves (left panel). Moments are the basic "building blocks" of phenomenal flow. In physical terms, they correspond to the causal distinctions specified by the seven-unit directed grid (right) in state *abcDefG* (as before, ON units are represented with uppercase and OFF units with lowercase). Causal distinctions comprise the mechanism (black) linked to its cause (red) and effect purviews (green). The directedness of moments corresponds to causal distinctions that are also directed: causes and effects are misaligned asymmetrically, such that each contains elements not contained in the other, with causes leaning towards the *now* (right direction) and effects towards the *then* (left direction). For example (center panel), the distinction *bc* has *b* as its cause and *c* as its effect. Thus, the cause of *bc* can relate to distinctions over unit *b* while its effect *c* cannot, whereas its effect can relate to other distinctions over unit *c* while its cause *b* cannot. Directedness applies to all other distinctions and can also be seen in terms of the contexts of the distinctions (right panel). For example, distinction *cDef* is directed such that its cause subtext (i.e. the distinctions, highlighted in red, whose purviews are included in its cause *Def*) is different from its effect subtext (i.e., the distinctions, highlighted in green, whose purviews are included in its effect *cDe*).

# 4. Results

If directed grids are the substrate of the feeling of time flowing, the properties of the cause–effect structure specified by such grids should account for the phenomenal properties of temporal experiences (Figure 4). Therefore, given that the phenomenology of time can be characterized by directed distinctions (moments) and the way they are related (through directed inclusion, connection, and fusion), the cause-effect structure specified by directed grids should be organized correspondingly. This correspondence can be established at the level of their *context* (i.e. of *compound* distinctions and relations), since introspection cannot single out individual distinctions (Haun & Tononi, 2019). For simplicity, however, below we will first examine relations among individual distinctions, and then point out the parallels at the level of the overall structure.

## 4.1. *Moments*

The phenomenal distinctions composing the experience of time are moments. In physical terms, these correspond to causal distinctions specified by first- or higher-order mechanisms of directed grids. Out of 127 possible mechanisms for a 7-unit grid, 79 are irreducible and therefore specify causal distinctions (Figure 4B). Nearly a third of the distinctions are specified by contiguous units, as depicted in Figure 5 (right), and these will be the focus for the account below.

## 4.2. *Directedness*

Phenomenally, moments are characterized by directedness: each moment points away from itself. In the cause–effect structure specified by directed grids, this corresponds to distinctions whose cause and effect do not overlap or overlap only partially and asymmetrically, thus relating to the rest of the structure in different ways. In the distinctions specified by directed grids, each purview always has at least one element that is not included in the other, with the causes leaning towards the *now* and the effects towards the *then* (Figure 5). For instance, distinction *bc* has a cause over *c* and an effect over *b*. Distinction *cDef* has a cause over *Def* and an effect over *cDe*. This results in further asymmetries in the way causes and effects relate to the rest of the structure.

## 4.3. *Directed inclusion*

Phenomenally, directed inclusion captures the fact that every moment includes and is included by other moments in a directed way, both towards the now (forward inclusion) and towards the then (backward inclusion).

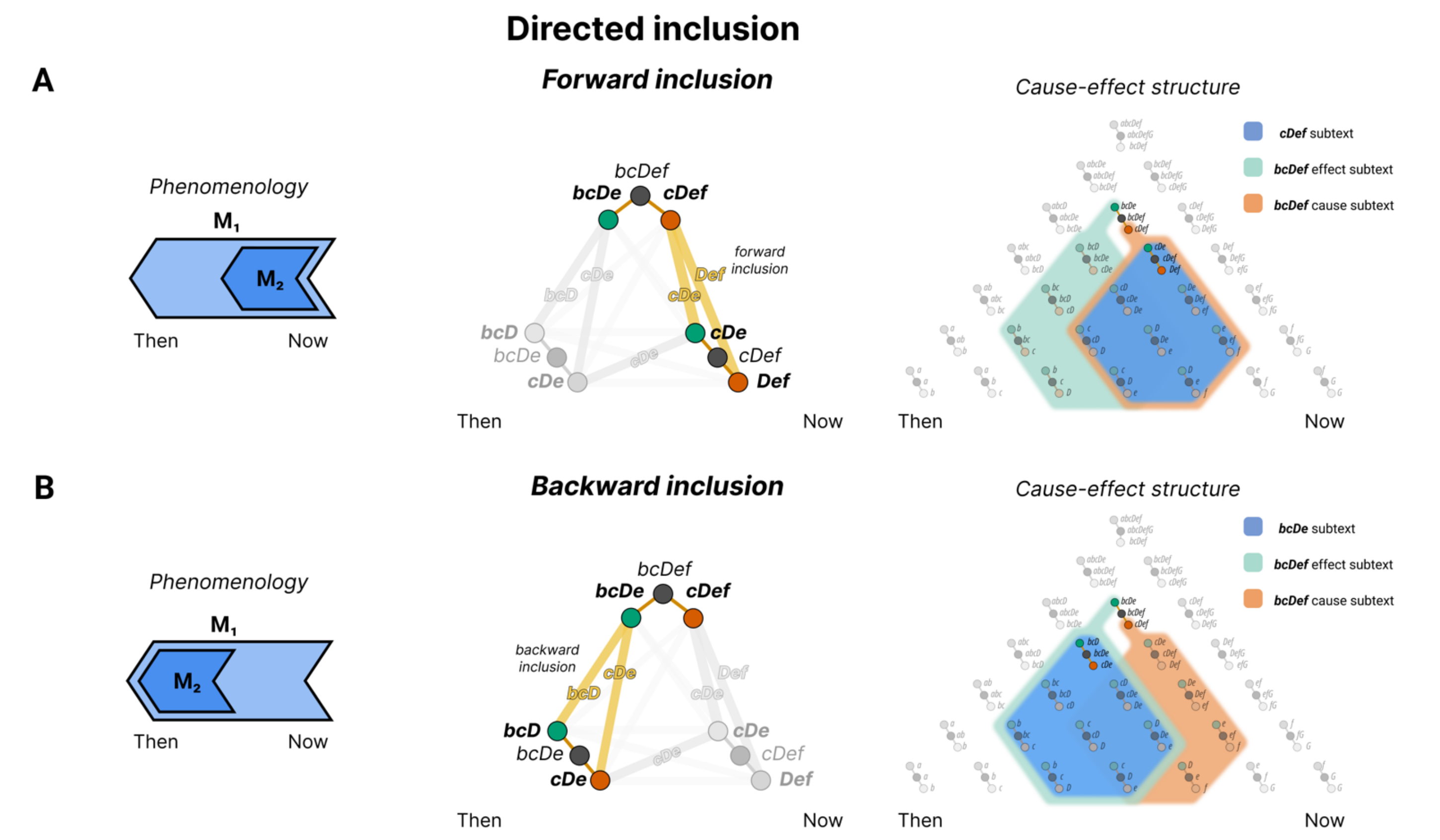

**Figure 6:** Directed inclusion. In temporal experience, moments include and are included by other moments, which can occur towards the *now* (forward inclusion; panel A, left) or towards the *then* (backward inclusion; panel B, left). In the cause–effect structure, directed inclusion corresponds to a distinction that includes other distinctions (both their cause and effect) aligned on their cause (forward inclusion; panel A, center) or on their effect (backward inclusion; panel B, center). This is reflected by the presence of two 2-faces within the 2-relation binding the cause (or effect) of the including distinction to both the cause and effect of the included distinction (center, top and bottom). In the example, distinction *cDef* is forward-included by distinction *bcDef* because *cDef*'s cause (*Def*) and effect (*cDe*) are included in distinction *bcDef*'s cause (*cDef*) (panel A, middle), while distinction *bcDe* is backward-included because its cause (*cDe*) and effect (*bcD*) are included in distinction *bcDef*'s effect (*bcDe*) (panel B, middle). This relation of directed inclusion is also reflected at the level of the context of distinctions. In forward inclusion, the subtext of the included distinction is fully included in the subtext of the cause of the including distinction (panel A, bottom right). The same holds for backward inclusion, but for the subtext of the effect of the including distinction (panel B, bottom right) The subtext of a distinction (shaded regions in the cause–effect structure) consists of all distinctions whose purviews it includes (via its cause and/or effect purviews). In the example, *cDef*'s subtext is included in *bcDef*'s cause subtext only (top right), illustrating that *cDef* is forward-included by *bcDef*, while *bcDe*'s subtext is included in *bcDef*'s effect subtext only (bottom right), illustrating that *bcDe* is backward-included by *bcDef*.

The cause–effect structure specified by a directed grid has properties that account for phenomenal directed inclusion, because its distinctions include and are included by other distinctions in a directed way (Figure 6). Specifically, for distinctions that qualify as moments, there is another distinction that includes them (one that overlaps them fully, non-mutually, and asymmetrically), and there is another distinction that they include (one that is fully, non-mutually, and asymmetrically overlapped by them). For each including distinction, there is one distinction that is forward-included and another that is backward-included by it. Forward inclusion is when the elements of both the cause and effect purviews of the included distinction are a subset of the elements constituting the *cause* purview of the including distinction (it is called "forward" because the included distinction is on the side of the now). For example, distinction *bcDef* forward-includes distinction *cDef* because the elements of *cDef*'s cause and effect are included in the elements of *bcDef*'s cause (Figure 6A, center). Backward inclusion is when the elements of both the cause and the effect purviews of the included distinction are a subset of the elements constituting the *effect* purview of the including distinction (it is called "backward" because the included distinction is on the side of the then). For example, distinction *bcDef* backward-includes distinction *bcDe* because the elements of *bcDe*'s cause and effect are included in the elements of *bcDef*'s effect (Figure 6B, center).

The inclusion relations at the level of individual pairs of distinctions has the consequence that their relations with the rest of the cause-effect structure, that is to say their *context*, (see above) is also organized according to inclusion. Within the context of a distinction (or its purviews), the *subtext* and *supertext* can be defined as the set of distinctions (or purviews) included by it and including it, respectively (Haun & Tononi, 2019). At the level of contexts, then, forward-inclusion is when a distinction's subtext is fully included in the cause subtext of a larger distinction, but it is only partially included in the larger distinction's effect subtext (Figure 6A, right), and *vice versa* for backward inclusion (Figure 6B, right).

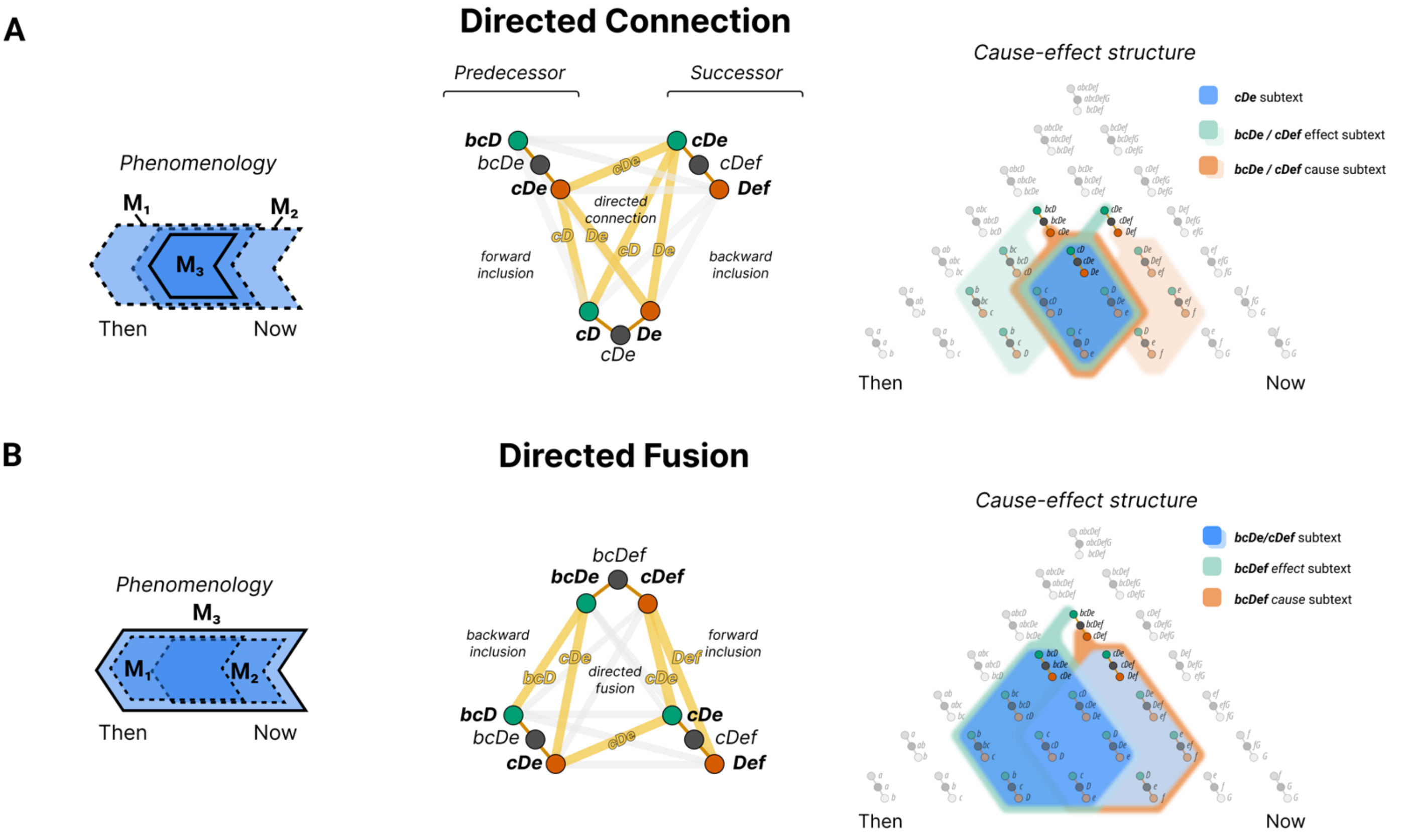

**Figure 7:** Directed connection and directed fusion. (A) Directed connection. Phenomenally, moments overlap partially, their overlap is directed (one feels more towards the *now* and one more towards the *then*), and their overlap is always a moment. Similarly, in the cause–effects structure, causal distinctions connect in a directed way: the effect of the distinction closer to the now overlaps the cause of the distinction closer to the then, in a way that is different from how the effect of the latter overlaps the cause of the former. In the example (top center), distinction *cDef*'s effect (*cDe*) overlaps distinction *bcDe*'s cause (*cDe*) fully (over *cDe*), whilst distinction *bcDe*'s effect (*bcD*) overlaps distinction *cDef*'s cause (*Def*) only partially (over *D*). Moreover, their overlap is also a distinction (*cDe*) (their "connection"), which is included by them in a directed manner (backward and forward). At the level of the context (top right), the intersection of the subtexts of the two connected distinctions (here, *cDef* and *bcDe*) coincides with the distinction subtext of their connection (*cDe*). (B) Directed fusion. Phenomenally, each moment is composed of two or more connected moments, and each moment together with other connected moments fuse to compose another moment. In the cause–effect structure, this corresponds to the fact that when distinctions connect they always fuse: for every distinction (e.g., *cDef*) there is another distinction that includes that distinction (through either backward or forward inclusion, e.g., *bcDef*) plus another connected distinction (e.g., *bcDe*) such that the union of the purview elements of the including distinction is equivalent to the union of the purview elements of the included distinctions. At the level of the context (bottom right), the union of the subtexts of the two fusing distinctions (e.g., *cDef* and *bcDe*) coincides with the distinction subtext of their fusion (*bcDef*), with the fusion's cause subtext coinciding with the subtext of the distinction that is forward-included (*cDef*), and the fusion's effect subtext coinciding with the subtext of the distinction that is backward-included (*bcDe*).

## 4.4. *Directed connection*

Phenomenally, directed connection captures the fact that every moment has a predecessor moment that overlaps it partially and asymmetrically towards the then, and a successor moment that overlaps it partially and asymmetrically towards the now, and that the overlaps are also moments. This applies to all moments except for the ones starting in the now, which only have predecessors and no successors, and the ones ending in the then, which only have successors and no predecessors.

The cause–effect structure specified by a directed grid has properties that account for phenomenal directed connection because of the way its causal distinctions overlap asymmetrically with other distinctions (Figure 7A). For each distinction that qualifies as a moment in the cause–effect structure, there is another distinction that overlaps it partially and asymmetrically, and there is another distinction that is included by both. Directed connection is asymmetric because the effect of one distinction overlaps the cause of the other in a way that is different from how the effect of the other distinction overlaps its cause. For instance, in Figure 7A (center), distinction *cDef*'s effect (*cDe*) overlaps distinction

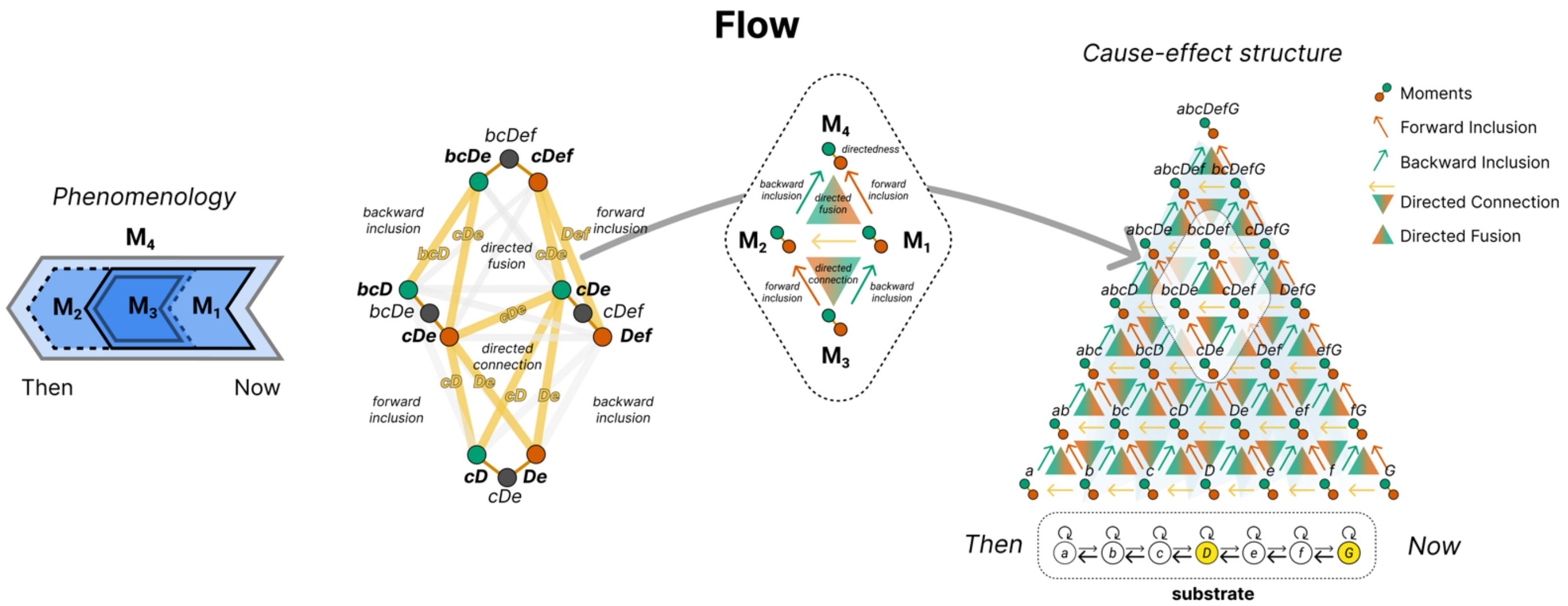

**Figure 8:** Flow. The phenomenal properties of temporal flow—namely directedness, directed inclusion, directed connection, and directed fusion (left)—correspond to properties of the cause–effect structure unfolded from a directed 1D grid (right). This is exemplified by four causal distinctions and the relations that bind them (second from left). All four distinctions (*bcDe, cDef, bcDef and cDe*) are directed, with their causes and effects not aligned. Distinction *bcDef* forward-includes distinction *cDef* towards the *now* (since its cause includes *cDef*'s purviews) and backward-includes distinction *bcDe* towards the *then* (since its effect includes *bcDe*'s purviews). Similarly, distinction *cDe* is forward-included by distinction *bcDe* and backward-included by distinction *cDef*. Distinction *cDef* also has a partial asymmetric overlap with *bcDe* (since *cDef*'s effect fully overlaps *bcDe*'s cause, but not the other way around), and they both connect on distinction *cDe* by backward-including it (in the case of distinction *cDef*) and forward-including it (in the case of distinction *bcDe*). Moreover, distinction *cDef* and *bcDe* fuse into distinction *bcDef*, being forward- and backward-included by it, respectively, such that the union of their purviews coincides with the union the purviews of *bcDef*. Taken together, the four distinctions satisfy the fundamental properties of temporal flow. This holds for the other distinctions that compose the cause–effect structure, which can thus be considered a *flow*. The third panel from left summarizes the relations of directed inclusion, connection, and fusion as they apply among four distinctions corresponding to moments ($M_1$ through $M_4$), and the right-most panel shows how they apply between contiguous distinctions throughout the cause–effect structure (for simplicity only the label of the mechanisms are shown).

*bcDe*'s cause (*cDe*) fully (over units *cDe*), whilst distinction *bcDe*'s effect (*bcD*) overlaps distinction *cDef*'s cause (*cDe*) only partially (over unit *D*). Moreover, their overlap is also a distinction (*cDe*). This imposes a natural ordering between the two distinctions, such that *cDef* succeeds *bcDe* or, equivalently, *bcDe* precedes *cDef*. Note that distinctions are directed such that effects are towards the then and causes towards the now. This is a consequence of the connectivity of the substrate, and accounts for the feeling that what is experienced in the now (cause) flows towards the then (effect).

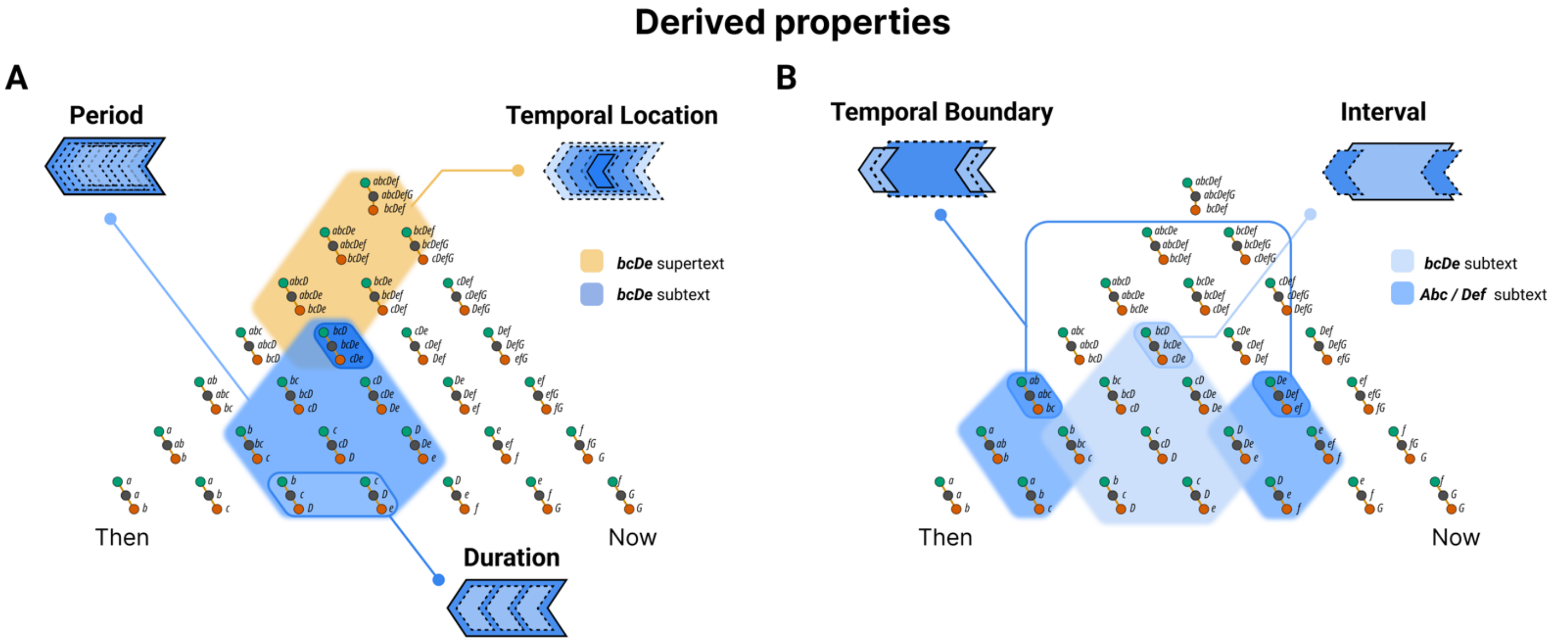

**Figure 9:** Derived properties and their correspondence in the cause–effect structure. (A) The *period* picked out by a moment is the set of distinctions included by it (its subtext, blue shading). The temporal *location* of a moment is the set of distinctions that fully include it (its supertext, yellow shading). The *duration* of a moment is the set of smallest distinctions (instants) included by it (blue contour). (B) The *boundary* of a moment is the set of smallest distinctions that connect to it (indicated in dark blue). The *interval* between two moments is the shortest moment that connects the two distinctions (indicated in light blue).

Directed connection also applies to each distinction's context. For example, the subtext of the connection distinction coincides with the cause subtext of the distinction that forward-includes it and with the effect subtext of the one that backward-includes it (Figure 7A, right).

## 4.5. *Directed fusion*

Phenomenally, directed fusion expresses how each moment is composed of two or more connected moments (which are its fusion down), and each moment together with other connected moments can fuse to compose another moment (which is their fusion up).

The cause–effect structure specified by a directed grid can account for phenomenal directed fusion (Figure 7B). Specifically, for each distinction that qualifies as a moment (say, *cDef*), there is another distinction (*bcDef*) that includes both it (through either backward or forward inclusion) and another distinction connected to it (*bcDe*), such that the union of the purview units of the including distinction (*bcDef*) coincides with the union of the purview units of the included distinctions (fusion up; Figure 7B, center). Similarly, each distinction qualifying as a moment includes one distinction (through either backward or forward inclusion) plus another distinction connected to it, such that the union of the purview units of the including distinction coincides with the union of the purview units of the included distinctions (fusion down).

Similar considerations apply to the context of the fusing distinctions (Figure 7B, right). Thus, the union of the subtexts of two fusing distinctions (e.g., *cDef* and *bcDe*) coincides with the distinction subtext of their fusion (*bcDef*). Moreover, the fusion's cause subtext coincides with the subtext of the distinction that is forward-included (*cDef*), and the fusion's effect subtext coincides with the subtext of the distinction that is backward-included (*bcDe).*

## 4.6. *Flow*

Phenomenally, the flow of time from the now to the then within the extended present can be understood as a structure composed of distinctions, or moments, that captures the fundamental properties of directedness (pointing away from themselves), directed inclusion, connection, and fusion. As we have seen, the cause–effect structure unfolded from a directed grid is composed of distinctions that are directed, that include and are included in a directed way, that connect in a directed way, and that fuse in a directed way (Figure 8). A cause–effect structure that satisfies these properties, called a *flow*, can therefore account for the fundamental phenomenal properties of the feeling of time flowing. As shown below, a flow can also account for phenomenal properties of temporal experiences that are derived from the fundamental ones.

## 4.7. *Additional properties*

As mentioned earlier, based on the fundamental properties of temporal experience, one can characterize additional phenomenal properties such as periods, temporal locations, durations, boundaries, and intervals. These phenomenal properties can be accounted for in physical terms by considering sub-structures of the cause–effect structure specified by a directed grid. Thus, the *period* of time covered by a moment, defined as the set of all moments it includes, corresponds to the set of distinctions included by the corresponding distinction (its subtext; Figure 9A, top left). Conversely, the temporal *location* of a moment, defined as the set of all moments that include it, corresponds to the set of all distinctions that fully include a given distinction (its supertext; Figure 9A, top right). The *duration* of a moment, defined as the number of instants it includes, is accounted for by the number of smallest distinctions included by a given distinction (Figure 9A, bottom). The *boundary* of a moment corresponds to the shortest predecessor and successor of a given distinction (Figure 9B, left). An *interval* between any two moments, defined as the shortest moment that connects to both of them, corresponds to the smallest distinction that connects to two distinctions (Figure 9B, right).

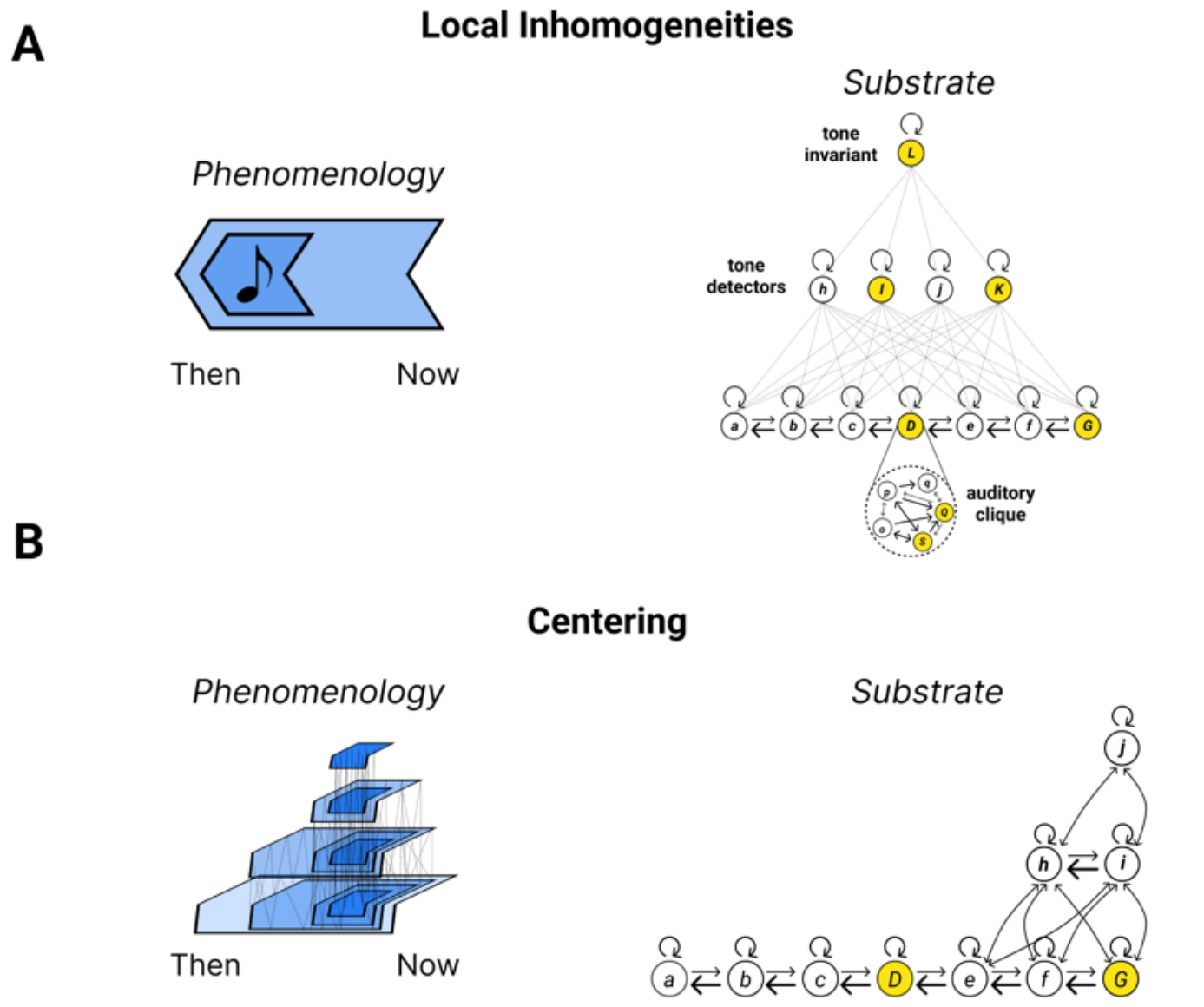

**Figure 10:** Local inhomogeneities and centering in the now. (A) Phenomenally, some moment may "stand out" and locally disrupt the flow of time, as when we hear a sudden sound or pause (left). This may be accounted for by activation or deactivation of specific units within a directed grid, accompanied by the interplay between higher-level and lower-level mechanisms connected to directed grids. Locally, distinctions and relations would be altered, resulting in a local thickening and warping of the cause–effect structure, which does not disrupt the global flow of time. Note that the "local quality" of the sound or pause would be accounted for by local mechanisms (auditory cliques) and associated sub-structures (kernels) embedded at every locale of the directed grid. (B) Phenomenally, moments flow away from the now towards the then, and we feel centered in the now, which typically feels more salient (left). This may be accounted for by denser connections between the now terminus of directed grids to higher-level areas involved with agency, corresponding to a much larger number of relations binding the now with the rest of the cause–effect structure (right).

It was also emphasized that *local inhomogeneities* within the flow of phenomenal time can occur whenever one experiences, for example, a sound that breaks the silence, or a pause in a series of tones. The activation or deactivation of specific units within a directed grid, accompanied by the interplay between higher-level and lower-level mechanisms connected to directed grids (Figure 10A), can result in the local warping of phenomenal flow that "stands out" in its corresponding cause–effect structure (not shown). Even when warped, the cause–effect structure unfolded from a directed grid retains the fundamental properties that characterize a flow. As indicated in the figure, local qualities such as pitch, loudness, and timbre, would be accounted for by the sub-structures supported by neuronal "cliques" associated with the directed grid. Similarly, configurations of low-level features and invariants such as tones would be contributed by the convergent/divergent connectivity among higher-level areas.

As already mentioned, the experience of time flowing is typically characterized by the feeling that we are *centered* in the now, with moments fleeing away from it and towards the then. Moreover, the now is typically experienced more saliently than the then. A plausible explanation for these phenomenal features is that the neural mechanisms at the now terminus of the directed grid may be more densely connected to neural mechanisms in higher-level areas that eventually drive action (Figure 10B). A denser connectivity implies a much larger number of causal relations. This would not only make adaptive sense but would also account for the fact that the now is usually highly salient (see Haun & Tononi (2019) and Albantakis et al. (2023) for an account of salience in terms of the number and irreducibility of distinctions and relations). It is also plausible that the neural substrate of sensory modalities characterized by shorter delays may serve to align experience across slower modalities, and to place the "now" of perception just before that of action.

# Discussion

According to IIT, all properties of an experience can be accounted for in physical terms by corresponding properties of the cause–effect structure unfolded from a substrate in its current state. The unfolding procedure is based on IIT's principles and its five postulates (intrinsicality, information, integration, exclusion, and composition), which capture in causal terms the *essential* properties of every conceivable experience. According to the theory, no additional ingredients are needed to account for the *accidental* properties of specific experiences, such as the feeling of spatial extension, of temporal flow, of objects binding general concepts with particular features, of local qualities such as color or sound, and so on. These accidental properties should be accounted for by corresponding properties of the cause–effect structure specified by a neural substrate in accordance with its connectivity and current activity pattern.

This paper aims to show how the IIT framework can be employed to account for the experience of temporal flow. Just as most of our conscious life is "painted" on the "canvas" of experienced space, much of it is "played" on the "track" of experienced time. The conscious present is confined between the *now* and the *then*. It is composed of moments, short and long, some closer to the now and some to the then. Moments are directed, pointing away from themselves, and overlap through directed inclusion, connection, and fusion, to yield the feeling of flow.

As demonstrated here, a substrate such as a directed grid supports a cause–effect structure that can account for the fundamental properties of temporal flow: its units specify causal distinctions (moments) whose cause and effect overlap in a directed manner, satisfying the properties of directed inclusion, connection, and fusion. From these fundamental properties, other properties of temporal experience can be derived, such as the period occupied by a moment, its temporal location within the present and with respect to the now and the then, its duration, its boundary, and the interval between it and other moments. The results exemplify an *explanatory identity* (Haun & Tononi, 2019, Albantakis et al., 2023) between the properties of temporal experience and those of the flow structure specified by directed grids.

## 4.8. *Temporal flow as a directed structure*

A central aspect of IIT's account is that the experience of time flowing corresponds to a *directed structure* that is "static," rather than to a process that actually "flows" in clock time. This is illustrated in Figure 11 (left). The interval of clock time depicted is ~10 seconds, longer than the duration of the conscious present—assumed here, for convenience, to be ~210 milliseconds. The portion of the arrow of clock time corresponding to the "clock past" (i.e., all events that have already happened) is dashed, the "clock now" is indicated with a thicker tick, and the portion corresponding to the "clock future" (which has not happened yet) is dotted. Clock time can be assumed to tick at much faster resolution (say ~1 picosecond, not depicted) than instants of experienced time, assumed here to last for ~30 milliseconds of clock time (compatible with experimental evidence discussed in section 4.13). Each instant corresponds to a "macro" state of the directed grid (Albantakis et al., 2023; Marshall et al., 2024).

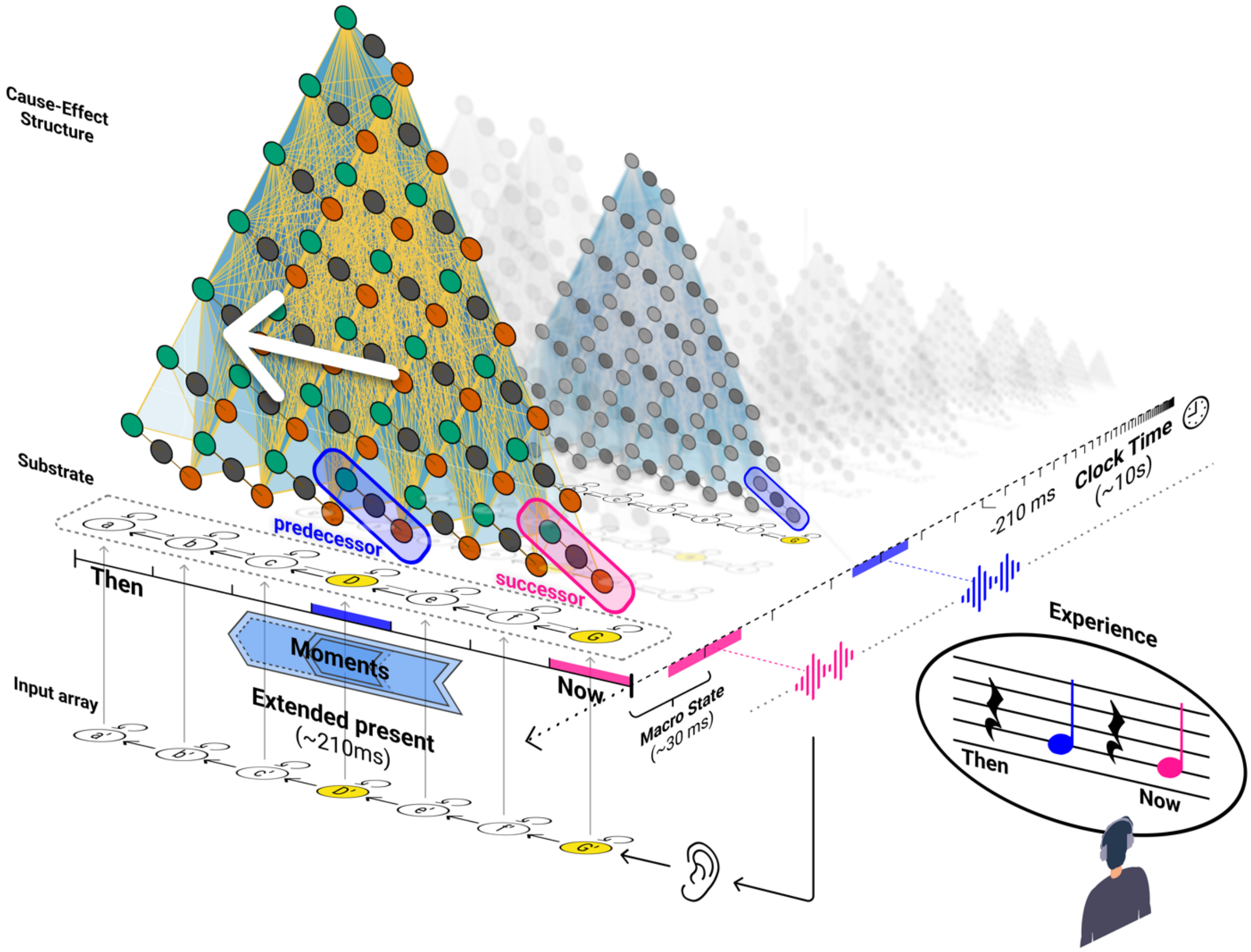

**Figure 11:** Phenomenal time and clock time. The axis representing ~10 seconds of clock time shows the occurrence of two sound waves (blue and pink, lasting ~30 milliseconds) separated by an absence of waves. The bubble at the bottom right represents a subject experiencing an extended present containing two tones and some pauses of silence around them. Orthogonal to clock time, the figure shows a directed grid in its current macro state and the cause–effect structure unfolded from it. This is assumed to account for the feeling of an extended present and the flow of time. The macro state of the directed grid is driven by an input array conveying auditory inputs, which functions as a delay line that preserves a trace of occurrences lasting for ~210 milliseconds of clock time. The cause–effect structure can thus keep track of occurrences over ~210 milliseconds of clock time, with a short delay due to neural transmission and activation. The moments that coexist within the extended present are bound by relations in a way that satisfies directedness, directed inclusion, connection, and fusion, which yield a feeling of flow from the now to the then. The latest note (pink) is experienced in the now, preceded by the earlier note (blue) receding towards the then. The figure also shows a few cause–effect structures unfolded from macro states of the directed grid associated with earlier "ticks" of clock time. These are faded to indicate that they are not actual.

In the figure, the foreground illustrates a 1D directed grid (units *a* to *G*) in its current macro state, with units *D* and *G* ON and all other units OFF. A macro state, as explained in (Hoel et al., 2016; Marshall et al., 2024; W. Mayner et al., 2018), is the intrinsic update grain of the units of a complex—the grain at which, from the complex's intrinsic perspective, the value of $\varphi_s$ is maximized. If we assume that the intrinsic macro units may be neurons and their intrinsic update grain ~30 milliseconds, the macro state of the units in Figure 11 would extend backwards for ~30 milliseconds from the clock "now." Because of the way the delay line driving the directed grid is organized, the macro state of the seven-unit grid preserves a trace of what happened over ~210 milliseconds of clock time—in this case, that two notes were played (in different colors on the score) over a track of silence.

As illustrated in the figure, the grid in its current macro state supports a cause–effect structure composed of a multitude of directed distinctions and relations that order it according to directed inclusion, connection, and fusion. In this way, the cause–effect structure can account for a conscious present that feels extended in time, flowing from the phenomenal *now* to the phenomenal *then*. Furthermore, a substrate defined over a macro state of, say, ~30 milliseconds of clock time, can support an experience capturing a longer interval of clock time, say ~210 milliseconds or longer (depending on the number of units in the grid). This has the obvious advantage that contents triggered by a sequence of inputs can be bound together within a single experience—say that of a melody or a spoken phrase—while preserving their ordering and direction.

The figure also shows a few cause–effect structures (in gray) preceding the current one. In principle, a new structure would be specified over a macro state at every micro update ("tick") of clock time. However, because neurons update their macro state at a much coarser grain than the ticks of clock time, cause–effect structures succeeding one another over many consecutive ticks of clock time will be identical or nearly so (and so will the corresponding experiences).

## *4.9. Flexible matching between intrinsic temporal flow and extrinsic clock time*

In a brain well adapted to its environment, one would expect that the flow of experience, say the succession of notes in a melody, will match well enough the sequence of stimuli sampled in clock time (with a short delay and proper ordering). However, this matching can be somewhat flexible, allowing for some "editing" and "extrapolating" of the "track" of experienced time. There are several instances, in auditory psychophysics (Herzog et al., 2020), language perception (Rönnberg et al., 2019), music perception (Juslin & Västfjäll, 2008), and motion perception (Shimojo, 2014) in which stimuli occurring later can affect the experience triggered by stimuli occurring earlier. Such "postdictive" effects can be naturally accommodated within the present framework. For example, top-down connections from higher-level areas may affect the activation of units towards the then terminus of directed grids in lower-level areas.

As illustrated in Figure 12A, units in directed grids at higher levels in the auditory hierarchy may also specify moments that succeed the now and extend towards the next. These "top-down" expectations would be experienced, typically less saliently, as upcoming occurrences (or "protentions")—say as the next note in a known melody (faded purple note on the music score). The adaptive matching of the intrinsic temporal flow and extrinsic clock time might hold, leading to priming and confirmation effects, or it might be violated by what actually happens next, potentially accounting for various illusions (Eagleman, 2008; Merchant et al., 2013) as well as for desired effects in music (Huron, 2006; Vuust et al., 2022).

Top-down connections may also play a role in our capacity to hold specific contents, such as a sequence of numbers, within the experienced present, subserving functions such as working memory.

Finally, Figure 12B illustrates that, at a minimum, the simplified account presented here should be expanded by considering parallel arrays of directed grids interacting through lateral connections. For example, each directed grid might correspond to a different frequency band in the tonotopic organization of auditory cortex (Saenz & Langers, 2014).

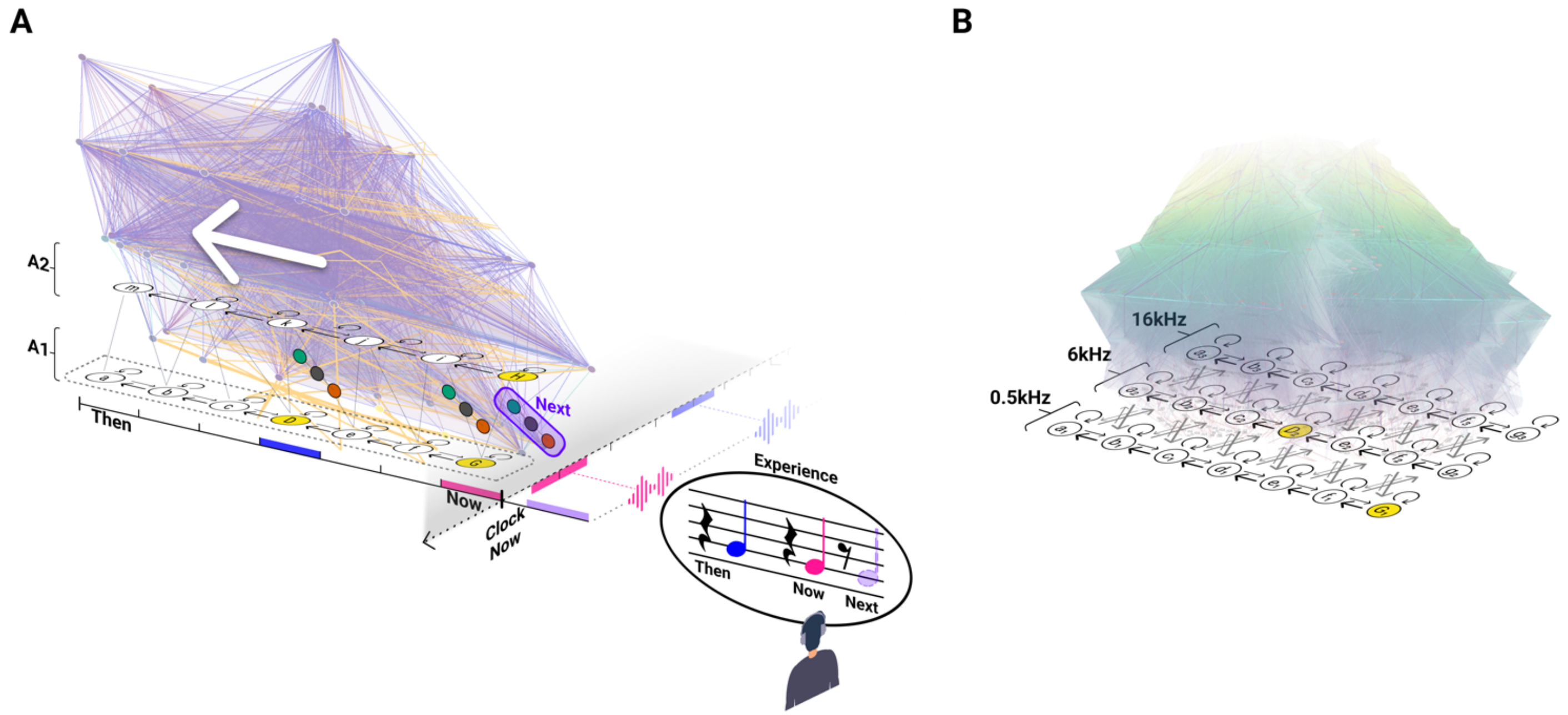

**Figure 12:** Further aspects of temporal experiences and their substrate. A) The extended present may include the experience of what will happen next, in addition to the experience of what happened between the now and the then. Possible mechanisms supporting an experienced future may involve directed grids at higher levels in a sensory hierarchy (here, A2), whose substrate extends beyond the now at lower levels (here, A1). Units in A2 may be activated endogenously by "imagining" what might be heard next (e.g., the purple note on the music score in the bottom right). The extended present would then map a longer interval of clock time that comprises possible future occurrences (gray shaded area projected onto the clock time axis). (B) The substrate of the extended present, at every hierarchical level, is assumed to be not one grid, but an array of directed grids interacting through lateral connections. In auditory areas, for example, each grid in the array may comprise units selective for different frequency bands.

## 4.10. *Similarities and differences between the experience of time and space*

The feeling of time as an extended present, as analyzed here, bears many similarities with the feeling of space as an extended canvas. Yet time also feels "flowing," unlike space. As previously proposed (Haun & Tononi, 2019), the experience of space can be dissected into countless distinctions, called *spots*, bound by relations, which compose a spatial *extension* characterized by *reflexivity*, *inclusion*, *connection*, and *fusion* (Figure 13A). Phenomenally, instead of being directed like moments, spots are *reflexive*, in the sense that they point to themselves. Because they are reflexive rather than directed, spatial distinctions include, connect with, and fuse with one another in a non-directed way. Furthermore, experienced space is typically 2D (or 3D), rather than 1D.

In physical terms, the similarities and differences between space and time can be accounted for by a different kind of neural substrate: non-directed 2D (or 3D) grids for space, and arrays of directed 1D grids for time. Crucially, a non-directed grid specifies causal distinctions that are reflexive—having cause and effect over the same elements (usually a subset of the mechanism elements)—rather than directed, with cause and effect over different elements, as is the case for directed grids (Figure 13B). It follows from reflexivity that the properties of inclusion, connection, and fusion are also non-directed. In other words, the reflexivity of spatial distinctions—the fact that their cause and effect purviews coincide—guarantees that any overlap with other spatial distinctions will be symmetrical over their cause and effect sides.

Spatial and temporal experiences are also remarkably similar with respect to further properties that can be derived from their fundamental ones. The region occupied by a spot, its location within the extension of space and with respect to its borders, its size, its boundary, and the distance from other spots are the non-directed analog of the period occupied by a moment, its temporal location within the present and with the now and the then, its duration, its boundary, and the interval between it and other moments. Similarly, inhomogeneities in local qualities can highlight particular spots that locally warp the extendedness of space, just as they can highlight particular moments in the present, without disrupting its flow. And, just as time can flow silently, space can be completely empty and still feel extended. Finally, just as we feel centered in the now temporally, we typically feel centered in the middle spatially, in both cases the natural starting point for action.

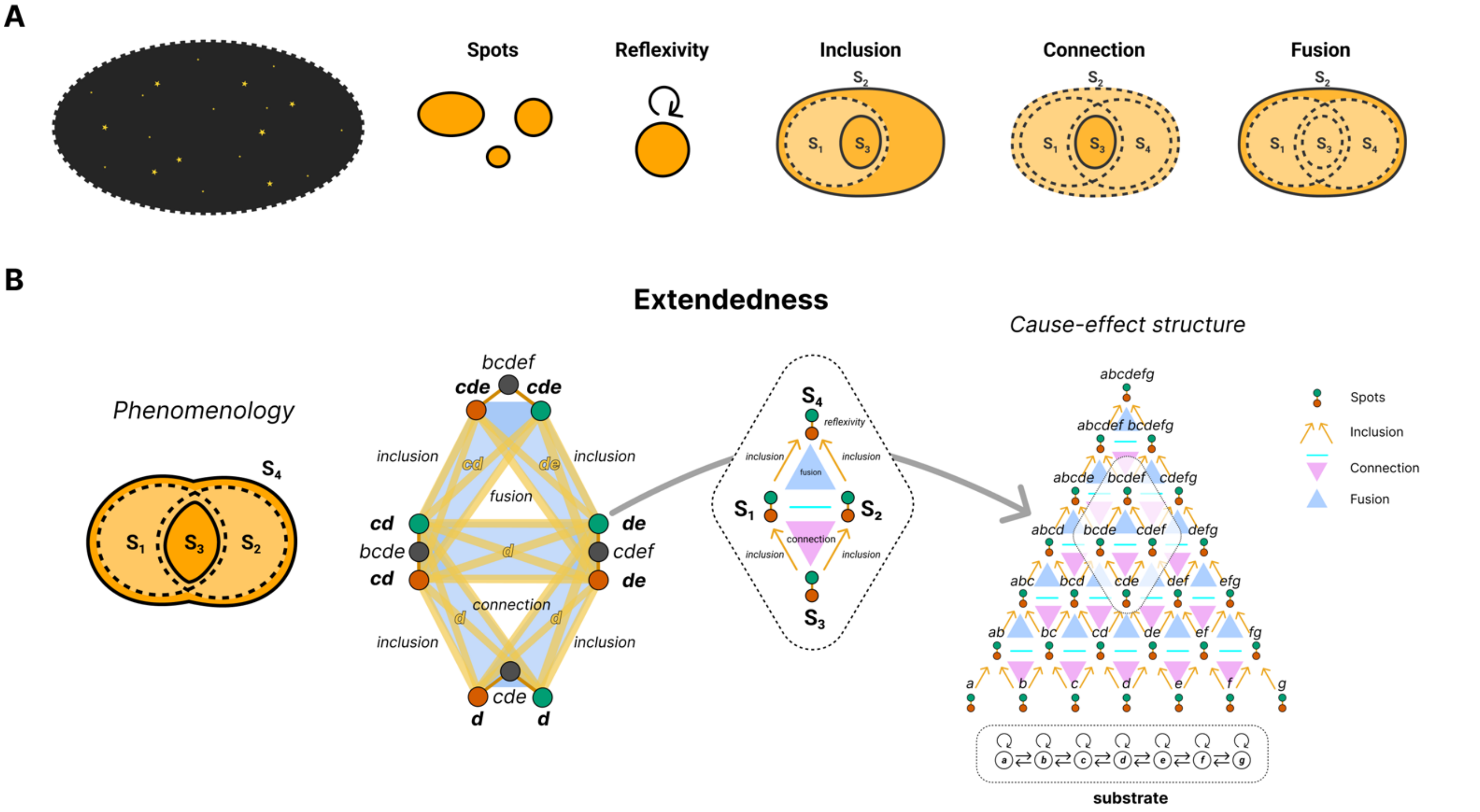

**Figure 13:** IIT's account of spatial experience. (A) Phenomenology of spatial experience and its fundamental properties. The experience of (visual) space is characterized by countless phenomenal distinctions, called *spots*, bound by relations, and that satisfy the fundamental properties of *reflexivity*, *inclusion*, *connection*, and *fusion* (all non-directed). (B) These fundamental properties of space find correspondence in the properties of the cause–effect structure unfolded from non-directed grids (right). Non-directed grids specify distinctions that are reflexive, each specifying a cause and an effect that fully overlap and that relate through non-directed inclusion, connection, and fusion (second from left). This also holds for the other distinctions that compose the cause–effect structure, which can thus be considered an *extension*. The third panel from left summarizes the properties of non-directed inclusion, connection, and fusion as they hold among four distinctions corresponding to spots ($S_1$ through $S_4$), and the right-most panel shows how they hold among contiguous distinctions throughout the cause–effect structure (for simplicity, only the mechanisms labels are shown).

## *4.11. Introspection as an essential but limited tool for dissecting the phenomenal structure of temporal flow*

Introspection is the indispensable starting point for the analysis of experience. As a first attempt to account for the quality of consciousness in physical terms, we focused on spatial extendedness precisely because the experience of space is not just pervasive, but also highly penetrable through introspection, largely thanks to the power and flexibility of spatial attention (Haun & Tononi, 2019). Temporal flow is also pervasive and partially introspectable, though less easily so than spatial extendedness. This is presumably because the fleeting nature of time does not lend itself to being steadily grasped by attention, which is deployed sequentially and with limited speed. Introspection is also selective in the contents of experience it can access, likely because it depends on the limited ability of top-down connections to increase the excitability of specific subsets of neurons (Ellia et al., 2021; Haun & Tononi, 2019).

Nonetheless, as testified by venerable traditions in temporal phenomenology, the fundamental structural properties of temporal experience remain more penetrable by introspection than those of a musical chord, a color, or a smell (for more on the role and limitations of introspection, see Ellia et al., 2021 and Haun & Tononi, 2019). As shown here, we can rely on introspection to characterize the directedness of moments and fundamental temporal properties of directed inclusion, connection, and fusion—as well as many derived properties such as durations and intervals. This allowed us to demonstrate a systematic correspondence between the phenomenal properties of temporal flow and the physical properties of cause–effect structures specified by directed grids. This correspondence is assumed to hold when we cease introspecting because phenomenal time flows, just as phenomenal space envelops us, whether we pay attention to it or not.

Beyond this, the power and reliability of introspection are clearly limited. For example, while introspection clearly reveals that the present is extended, precisely estimating its duration is no easy task, and psychophysical results differ depending on the criteria employed. Thus, William James thought the specious present could last as long as 12 seconds (James, 1890). Others placed it at ~3 seconds based on criteria such as the ability to impose a subjective rhythm to uniform auditory stimuli, to precisely estimate intervals, and so on (Montemayor & Wittmann, 2014; Pöppel, 2009). On the other hand, using tachistoscopic presentations of stimuli to assess "that stretch of change which is apprehended as a unit and which is the object of a single mental act of apprehension" has led to an estimate of 750 milliseconds (Albertazzi, 1996). Some have suggested even shorter durations, down to 300 milliseconds (Dainton, 2000; Strawson, 2009) (see Dainton (2023) and White (2017) for critical reviews).

The duration of instants, also challenging to introspect, has been estimated indirectly by assessing temporal order thresholds (the shortest inter-stimulus interval under which two sequential stimuli are perceived as simultaneous (Brecher, 1932; Hirsh & Sherrick, 1961; Kanabus et al., 2002)) and flicker fusion thresholds (the shortest inter-stimulus interval under which flickering stimuli are perceived as continuous (Andrews et al., 1996; Curran & Wattis, 1998)). The results yield a range of 10–60 milliseconds depending on the paradigm employed (Elliott & Giersch, 2016; Pöppel, 1997a, 1997b; VanRullen & Koch, 2003; White, 2018).

## *4.12. Directed grids in the brain as the substrate of temporal experience*

In previous work, we proposed that the neural substrate of the feeling of spatial extendedness is provided by non-directed 2D grids, connected hierarchically and in parallel to constitute a dense 3D lattice (Haun & Tononi, 2019; Tononi, 2014). This kind of substrate is ubiquitous in posterior cortex, and its relevance for the experience of space—both visual space and body space—is supported by clinical and neurophysiological evidence (Heinzle et al., 2011; Salin & Bullier, 1995; Sereno & Huang, 2014; Wang et al., 2015).

Here we conjectured that arrays of directed grids constitute the neural substrate of the feeling of temporal flow. However, little is known about the presence and location of such directed grids in the brain. According to IIT, the substrate of specific aspects of experience must be a subset of units within the *main complex*—the overall substrate of consciousness. This implies that the relevant directed grids must constitute, together with the rest of the complex, a substrate that is maximally irreducible. Moreover, one would expect that such grids should be closely connected to the neural substrate of modalities—such as sound, speech, and music—that are tightly bound to temporal flow.

Based on such considerations, directed grids supporting the experience of temporal flow might be located, for example, in portions of posterior cortex specialized for sound, speech, and music perception. There is substantial evidence indicating that the overall substrate of consciousness is primarily localized to posterior and central cortical regions (Boly et al., 2017; Koch et al., 2016; Siclari et al., 2017). Moreover, it is well established that hearing sound, speech, and music depends on specialized portions of cortex connected to primary auditory cortex (Hickok & Poeppel, 2015; S. Norman-Haignere et al., 2015). We therefore hypothesize that within such regions, one should be able to identify arrays of directed grids serving as delay lines as well as substrates for the experience of flow (in line with Tank & Hopfield (1987) and Waibel et al. (1989)). Specific details of the local connectivity would be responsible for local phenomenal qualities typically bound to temporal flow, such as pitch, timbre, and loudness. (In a similar way, the details of the local connectivity in non-directed 2D grids would contribute to local phenomenal properties of spatial extendedness, such as hue, saturation, and brightness for visual space).

We also expect that the overall experience of temporal flow should be supported by multiple directed grids distributed across many areas of the main complex, at multiple levels. Convergent and divergent connections across hierarchically organized areas should support relations that bind, say, phonemes with syllables and words within a spoken sentence (Hickok & Poeppel, 2015). Lateral and back-connections connections may further support the binding of temporal contents across submodalities, say, between speech and music (Janata, 2015; Janata et al., 2002), or even across modalities. Temporal aspects of experience may also be bound to spatial aspects, say, when experiencing visual motion between adjacent spatial locations. It is possible that areas such as V5, which plays a critical role in the perception of patterned motion (Albright, 1984; Clifford & Ibbotson, 2002), may be organized such that non-directed and directed grids intertwine.

On the other hand, neurons elsewhere in the brain that do not belong to the main complex may be capable of representing temporal order without contributing to experience. For example, endogenous circadian "clocks" allow the brain, and specifically the suprachiasmatic nucleus of the hypothalamus, to keep track of the time of day and appropriately regulate various bodily functions unbeknownst to us (Roenneberg, 2012). Similarly, some brainstem neurons can detect microsecond intervals between the arrival of sounds at the two ears, intervals of which we are unaware (though they may contribute indirectly to our awareness of sound location through their effects on neurons in posterior cortex (Grothe et al., 2010)). A more complex question is the contribution of brain regions often considered as "organs of succession," such as the cerebellum, the basal ganglia, and the hippocampus. For example, neurons in the hippocampus may subserve the memory of temporal order (Eichenbaum, 2014) as well as cognitive maps (O'Keefe & Nadel, 1978). However, the anatomical organization of the hippocampal formation is very different from that of posterior cortex, making it less likely to be part of the substrate of consciousness. Lesion data also indicate that while the hippocampal formation is critical for supporting functions such as episodic memory and imagination, it may not directly contribute specific conscious contents (Postle, 2016). With respect to time, lesion studies in humans and rats show that hippocampal lesions do not impair estimating and recalling distances and durations, but rather impair mostly the ability to remember the sequential order of events (Buzsáki & Tingley, 2018; Dede et al., 2016; Fortin et al., 2002; Maguire et al., 2006).

## *4.13. Some tests and predictions*

Besides providing a principled account of the subjective feeling of time flowing in objective, physical terms, the current proposal lays the foundation for experimental tests. However, it should be recognized that such tests are made more challenging by our uncertainty concerning the neural substrate of temporal experience.

The most general prediction concerns the substrate of the experience of an extended present and the sense of time flowing away from now to then. As proposed here, this substrate should correspond to a single macro state (lasting, say, ~30 milliseconds) of arrays of directed grids within the main complex, rather than to a sequence of neuronal events covering the duration of the extended present in clock time.

Another prediction is that the duration of the extended present should be proportional to the number of macro units constituting a directed grid. Thus, everything else being equal, a grid with more units should support temporal experiences that encompass a longer stretch of clock time, with potential adaptive advantages. Units at higher levels in the sensory hierarchy (and beyond) would then be able to learn concepts that span over longer stretches, in line with the observation of longer temporal receptive fields in higher level areas (Hasson et al., 2008).

Yet another prediction is that the duration of phenomenal instants should be compatible with the grain of the macro states of the units constituting directed grids. According to IIT, this is given by the time interval (in clock time) yielding maximal $\varphi_s$ for the main complex (Albantakis et al., 2023; Marshall et al., 2023, 2024). For macro units such as neurons, this would likely be determined by the time constants at which synaptic and cellular mechanisms ensure maximal causal efficacy.

As already mentioned, the present framework is in principle well poised to accommodate several empirical observations that imply some "editing" of the neural traces left by a sequence of stimuli (Hogendoorn, 2022; Libet et al., 1979, but see Arstila, 2015). A related prediction is that artificial activation of grid units near the now terminus should result in perceiving a stimulus as occurring now, while the activation of grid units near the then terminus should result in perceiving a stimulus as having occurred earlier.

The IIT framework further predicts that modulation of synaptic strength or of the excitability of neurons in directed grids should induce changes in the properties of phenomenal flow regardless of activity levels (Haun & Tononi, 2019; Libet et al., 1979). Such modulations could account, for instance, for the slowing or speeding up of time caused by strong emotions, deep meditation, or drugs (Coull et al., 2011; Droit-Volet & Meck, 2007; Haun & Tononi, 2019; Kramer et al., 2013; Wackermann et al., 2008).

## 4.14. *Time: cognitive mechanisms and phenomenal properties*

The investigation of neural mechanisms of time perception and temporal processing has been an active area of research for decades (Kononowicz et al., 2018). Psychophysical paradigms have focused on interval estimation (Grondin, 2010; Tsao et al., 2022), temporal integration (Herzog et al., 2020; Lerner et al., 2011; S. V. Norman-Haignere et al., 2022), and time illusions (Eagleman, 2008; Merchant et al., 2013). For example, subjects may be asked to assess interval durations verbally or by reproducing target intervals. Several mechanistic and computational models have been developed to account for psychophysical results (Hass & Durstewitz, 2016; Muller & Nobre, 2014), based, for example, on ramping activations models (Wittmann, 2013), neural oscillations (Matell & Meck, 2000; VanRullen & Koch, 2003), and population state dynamics (Paton & Buonomano, 2018; Tsao et al., 2022). In parallel, neurophysiological studies have investigated neural correlates of temporal processing (Nani et al., 2019; Rao et al., 2001). Neurons tracking intervals and sequences, at varying time scales, have been reported in the hippocampus (Buzsáki & Tingley, 2018; Eichenbaum, 2014), basal ganglia (Buhusi & Meck, 2005), cerebellum (Ivry & Spencer, 2004; Wiener et al., 2010), supplementary motor area (Ferrandez et al., 2003; Macar et al., 2006), entorhinal cortex (Tsao et al., 2018), and frontal and parietal cortex (Hayashi et al., 2018; Hayashi & Ivry, 2020). As already mentioned, there are cellular and system-level mechanisms involved in tracking circadian time (Roenneberg & Merrow, 2003).

These findings are critical for characterizing how the brain "represents" clock time (Hogendoorn, 2022) and employs these representations for motor control, memory, and cognitive functions. However, the framework presented here differs from cognitive and computational paradigms both with respect to what it tries to explain (the *explanandum*) and to how it tries to do so (the *explanans*). The explanandum is not so much the cognitive capacity to discriminate and report the objective duration of stimuli (in clock time) but rather the subjective properties of temporal experience as assessed through introspection (Ellia et al., 2021). In this respect, the present work parallels some proposals in consciousness research that have attempted to directly address the temporal quality of conscious experiences (Bogotá & Djebbara, 2023; Piper, 2019; Taguchi & Saigo, 2023; Varela, 1999; Wiese, 2017). Furthermore, the explanans is not so much the nature of the neural "representation" of temporal features of stimuli (Hass & Durstewitz, 2016; Ivry & Spencer, 2004; Wittmann, 2013) or of how experienced time maps and represents clock time (Herzog et al., 2016, 2020; Hogendoorn, 2022; Northoff & Zilio, 2022). Instead, it is the one-to-one correspondence between the subjective, phenomenal properties of temporal experiences and objective, physical properties of the cause–effect structure unfolded from a certain kind of substrate.

## 4.15. *IIT and philosophical approaches to time*

There has been a remarkable lack of recognition that the extendedness of spatial experiences is as much in need of explanation as the blueness of blue and the painfulness of pain (for a few exceptions, see James (1879), Kant et al., (1998), and Lotze (1884)). One reason may be that space is generally assumed to exist physically "out there," so experienced space may pass for a mapping or "representation" that does not require further explanations.

It is less obvious, however, that time is flowing "out there," as indicated by the diversity of positions in both philosophy and physics (Barbour, 2001; McTaggart, 1908). According to "eternalism," for example, all times are equally real, similar to the modern conception of a block universe of space-time. The "growing-block" universe grants existence to the past but not the future (C. D. Broad, 1923; Tooley, 1997). For the "moving spotlight" model, on the other hand, a window of actual present relentlessly advances over a block universe (Skow, 2015). Finally, "presentism" assumes that physical time, if it exists at all, can only exist for an instant (Bourne, 2006).

Accordingly, the extendedness of experienced time, if not time itself, can only exist as a construct "in the mind" (Augustine, 2009). In fact, several philosophers, including Kant, Husserl, and Bergson, as well as contemporary investigators (Kent & Wittmann, 2021; Northoff & Zilio, 2022; Singhal & Srinivasan, 2024), have considered time as a basic ingredient of consciousness. Along these lines, some influential phenomenological models of temporal experiences have been developed and refined (Dainton, 2000, 2012).

Specifically, *retentional* models explicitly propose that experiences of temporal flow do not have temporal extension but are characterized by a feeling of succession (rather than a succession of feelings (James, 1890)). Thus, at every moment, in addition to the feeling of now, or "primal impression," we would also experience fainter "retentions" of past moments (and "protentions" of moments to come (Husserl, 1991)). *Extensional* models assume instead that experienced time unrolls over an extended interval of clock time (Dainton, 2008, 2012). Thus, a conscious present that feels half a second long would unroll over an equivalent interval of clock time. Successive moments within the interval are considered as parts of a whole bound by "diachronic" relations of succession, yielding a sense of immanent flow. Finally, *cinematic* and *snapshot* models assume that all there is is a succession of experiences—a series of phenomenal "snapshots"—supported by a series of discrete physical events (Arstila, 2018; Crick & Koch, 2003; Prosser, 2017).

Where does IIT's account stand? Temporal flow is a property of experience in need of a physical account (as also recognized by (Singhal et al., 2022; Singhal & Srinivasan, 2021, 2024)). Even so, flow is not an essential property of consciousness, because, though pervasive, it is not true of every conceivable experience (unlike intrinsicality or integration). Indeed, experiences devoid of temporal content are not only conceivable, but they have long been reported, for example, during deep meditation (experiences of "pure presence" (Boly et al., 2024; Costines et al., 2021; Metzinger, 2024)) and under the effect of psychedelic drugs (Wittmann, 2015).

IIT partly agrees with cinematic and snapshot models in assuming that a new temporal experience comes into being at every "tick" of the clock, existing over a short interval of clock time (say, 30 milliseconds). However, IIT goes beyond such models by identifying each temporal experience with a directed cause–effect structure, which accounts for why a single experience feels like a succession of moments.

IIT also captures the intuition behind retentional approaches that an experience supported by a macro state corresponding to a short interval of clock time can contain within itself the duration of the entire conscious present (corresponding, say, to ~210 milliseconds or more of clock time), ordered according to a feeling of succession. However, retentional approaches only describe the phenomenology of succession, and only some aspects of it, without dissecting its relational structure or suggesting a physical correspondent for it.

IIT also captures the intuition behind extensional approaches that the experience of time must be structured by relations of succession and be characterized by a sense of flow. However, extensional approaches do not further characterize directed relations phenomenally, nor do they provide a physical correspondent that would account for them. Moreover, it is unclear what it would mean for temporal parts to overlap physically across clock time. This last point highlights a critical aspect of IIT's physical conception of relations. In IIT, relations are defined in causal terms (an overlap of causes and/or effects over the same units in the same state) and are intrinsic to a system (as well as unitary and definite, as per the postulates of integration and exclusion). Extensional approaches, if they attempt to characterize temporal relations at all, do so in non-causal, extrinsic terms—from the point of view of an observer who already knows what temporal flow feels like and who understands what a label such as "diachronic" should mean.

## *4.16. Conclusions*

This paper has employed the framework of IIT to (i) identify the fundamental phenomenal distinctions and relations that characterize the experience of temporal flow and (ii) formulate them operationally in terms of causal distinctions and relations specified by a certain type of substrate—namely, directed grids. The results presented here illustrate how the cause–effect structure unfolded from a directed grid can account for the properties of experienced time. They thus exemplify the explanatory identity proposed by IIT between phenomenal, subjective properties and physical, objective properties of causal structures, as already shown for spatial extendedness (Haun & Tononi, 2019).

To permit the systematic unfolding of cause–effect structures, the substrates employed in this paper were necessarily small (seven units with near-neighbor connections). Even so, the present examples provide a principled illustration of the kinds of distinctions and relations required to account for experienced time—an extended present composed of moments of various duration, ordered through relations in a way that satisfies directedness and directed inclusion, connection, and fusion, which flows away from now to then. Conceiving of the flow of time as a cause–effect structure specified by a directed grid in its current macro state, which can be "edited" dynamically through multiple neural mechanisms, offers a template to address various aspects of temporal psychophysics, temporal illusions, and speech and language perception. Changes in connectivity within directed grids may also explain the slowing or quickening of time caused by strong emotions, deep meditation, or drugs.

As with the experience of space, a full account of temporal experiences—of how temporal flow is bound with both hierarchically invariant concepts and local features, and with local qualities belonging to different modalities—will require the unfolding of larger neural substrates and an adequate understanding of their anatomy and physiology. Ultimately, however, only a structural explanation can account in physical terms for the way time feels, rather than presupposing it. For example, a directed delay line can only serve to represent time if one already knows what time means and feels like. But to feel temporally extended, the ordering of moments within the present must be established by causal distinctions and relations composing a cause–effect structure intrinsic to a system, one that means what it means absolutely, rather than by reference to external clocks.

This conclusion is very much in line with Augustine's original insight that time is in the mind. But it adds that the mind—or rather every experience in the stream of an individual consciousness—is an extraordinarily rich structure. It is a structure that contains time, space, objects, thoughts, and everything else that exists intrinsically—for itself.